\documentclass[twocolumn]{aastex6}

\usepackage{graphicx}
\usepackage[suffix=]{epstopdf}
\usepackage{natbib}
\usepackage{amsmath}
\usepackage{url}
\usepackage{xspace}

\providecommand{\e}[1]{\ensuremath{\times 10^{#1}}}

\newcommand{\Kepler}{\textsl{Kepler}\xspace}

\begin{document}
\title{The Kepler Catalog of Stellar Flares}

\shorttitle{Kepler Flare Catalog}
\shortauthors{Davenport}

\author{
	James R. A. Davenport\altaffilmark{1,2}
	}

\altaffiltext{1}{Department of Physics \& Astronomy, Western Washington University, Bellingham, WA 98225}
\altaffiltext{2}{NSF Astronomy and Astrophysics Postdoctoral Fellow}

\begin{abstract}
A homogeneous search for stellar flares has been performed using every available \Kepler light curve. An iterative light curve de-trending approach was used to filter out both astrophysical and systematic variability to detect flares. The flare recovery completeness has also been computed throughout each light curve using artificial flare injection tests, and the tools for this work have been made publicly available. The final sample contains 851,168 candidate flare events recovered above the 68\% completeness threshold, which were detected from 4041 stars, or 1.9\% of the stars in the \Kepler database. The average flare energy detected is $\sim$$10^{35}$ erg. The net fraction of flare stars increases with $g-i$ color, or decreasing stellar mass. For stars in this sample with previously measured rotation periods, the total relative flare luminosity is compared to the Rossby number. A tentative detection of flare activity saturation for low-mass stars with rapid rotation below a Rossby number of $\sim$0.03 is found. A power law decay in flare activity with Rossby number is found with a slope of -1,  shallower than typical measurements for X-ray activity decay with Rossby number.
\end{abstract}

\section{Introduction}

Flares occur on nearly all main sequence stars with outer convective envelopes as a generic result of magnetic reconnection \citep{pettersen1989}. These events occur stochastically, and are most frequently observed on low-mass stars with the deepest convective zones such as M dwarfs. Solar and stellar flares are believed to form via the same mechanism: a magnetic reconnection event that creates a beam of charged particles which impacts the stellar photosphere, generating rapid heating and the emission we observe at nearly all wavelengths. Numerical simulations are now able to describe much of the physics for solar and stellar flares and their effect on a star's atmosphere \citep{allred2015}.

Flare occurrence frequency and event energy are connected to the stellar surface magnetic field strength. Reconnection events on the Sun typically occur around a sunspot pair (or bipole) or between a group of spots. Surface magnetic field strength deceases over the life a star, due to a steady loss of angular momentum which quiets the internal dynamo \citep{skumanich1972}. Older, slowly rotating stars like our Sun exhibit smaller and fewer starspots, while young, rapidly rotating stars can produce starspots that are long lived and cover a significant portion of the stellar surface. Flares are known to follow this same basic trend \citep{ambartsumian1975}. For example, young T Tauri systems are known to be highly active with frequent flares \citep{haro1957}. Maximal flare energies have also been proposed as a means for constraining the age of field stars \citep[e.g.][]{parsamyan1976,parsamyan1995}.

The duration of a star's life that it produces frequent large spots and flares may dramatically affect planetary, atmospheric, and biological processes, and thus impact planet habitability. This is particularly important for planets around low-mass stars, whose flares can produce extremely high amounts of UV and X-ray flux, and whose active lifetimes are much longer than Solar-type stars \citep{west2008}. To better understand the impact flares might pose for habitability, \citet{segura2010} modeled the affect of a single large stellar flare on a Earth-like planet's atmosphere. For a single large flare this study found only a short timescale increase in biologically harmful UV surface flux, and full planetary atmosphere recovery within two years. However, due to the possibility of repeated flaring and constant quiescent UV emission, concerns remain about UV flux from active and flaring stars, and their impact on planetary atmosphere chemistry \citep{france2014}. Given the variety of possible exoplanetary system configurations, it may also be possible for stellar activity and planetary dynamics to conspire to improve planetary  habitability conditions \citep{luger2015}. While the impact flares have to planet habitability is an ongoing topic of research, they pose a clear difficulty in exoplanet detection and characterization \citep{poppenhaeger2015}.

Due to their short timescales and stochastic occurrences, generating a complete sample of flares for a single star has been very resource intensive, and has only been accomplished for a handful of active stars. Contrast between flares and the quiescent star is also greatest for cooler stars such as M dwarfs, and has led to fewer flare studies for field G dwarfs. Flare rates for ``inactive'' stars like the Sun are largely unconstrained. However, recent space-based planet hunting missions like \Kepler \citep{borucki2010} have started to collect some of the longest duration and most precise optical light curves to date. These unique datasets are ideal for developing complete surveys of stochastic events like flares from thousands of stars, and have begun to revolutionize the study of stellar flares. For example, \citet{davenport2014b} gathered the largest sample of flares for any single star besides the Sun using 11 months of \Kepler data, and used this homogeneous sample to develop an empirical template for single flare morphology. To help characterize the environments of planets found using \Kepler, \citet{armstrong2016} have investigated the rates of very large flares for 13 stars that host planets near their habitable zones. \citet{maehara2012} have used \Kepler data to show a connection between flare rate and stellar rotation in field G dwarfs, in general agreement with activity--age models.

In this paper I present the first automated search for stellar flares from the full \Kepler dataset. The flare event sample generated here is unique in carefully combining both long and short cadence data to accurately measure each star's flare rate over the entire \Kepler mission. I have also performed extensive flare injection tests for multiple portions of each light curve, quantifying the completeness limits for flare recovery over time. I demonstrate the utility of this large sample by comparing the flare activity level with stellar rotation and Rossby number, which reveals a clear connection between flares and the evolution of the stellar dynamo as stars age.

\section{{\it Kepler} Data}
\label{sec:data}

\Kepler is a space-based telescope, launched in 2009 as NASA's 10th Discovery-class mission, with the goal of constraining the rates of transiting Earth-like planets around Sun-like stars. Achieving this science goal required observing a single large field of view of 115 sq deg with a few parts-per-million photometric accuracy, monitoring $\sim$150,000 stars simultaneously with a fairly rapid cadence, and observing continuously for nearly 4 years. While the exoplanet yield has been wildly successful \citep[e.g.][]{jenkins2015}, \Kepler has been equally fruitful in studying the astrophysics of field stars. For the first time, asteroseismology with \Kepler has provided information on the internal structure of stars besides our Sun, which places powerful constraints on their masses, radii, and ages \citep{chaplin2010,chaplin2013}. \Kepler's precision light curves have also enabled stellar rotation to be characterized for tens of thousands of stars \citep{reinhold2013,mcquillan2014}, shedding new light on angular momentum and dynamo evolution.

The unique sample size, light curve duration, and photometric precision makes \Kepler an ideal platform for studying stellar flares. \citet{walkowicz2011} observed many K and M dwarfs with prominent flare events in the preliminary \Kepler data release, finding correlations between flare rates, spectral type (or temperature), and quiescent variability levels. Defining the rate of large energy ``superflares'' on Solar-type stars from \Kepler is an important aspect for characterizing exoplanet habitability and understanding the early life of the Sun \citep{maehara2015}. Flares have been observed across a wide range of spectral types with \Kepler \citep{balona2015}, and the details of flare morphology in these data are now an active area of research \citep[e.g.][]{davenport2014b, pugh2015}.

\Kepler observed targets using two cadence modes. The vast majority of stars were observed using the ``long'', 30-minute cadence mode, and were observed continuously for most of the \Kepler mission. A small number of targets were selected for ``short'', 1-minute cadence observations, often for only a fraction of the \Kepler mission. Most \Kepler flare studies to date have focused on the long cadence light curves, which provide the best data for complete samples of large energy events such as superflares. However, flare occurrence frequency is inversely proportional to the event energy, and short cadence data is critical for detecting smaller energy, shorter timescale events, as well as characterizing the temporal morphology of superflares.

For this study I analyzed every available long and short cadence light curve from the primary \Kepler mission, obtaining the most recently available version of the Quarter 0--17 light curves, known as Data Release 24. Light curves are stored as {\tt .fits} tables that contain both the Simple Aperture Photometry (SAP) data, as well as the Pre-search Data Conditioning (PDC) de-trended data. Since the PDC light curve de-trending can be affected by the flares being searched for, the SAP light curves were used instead, as was done in \citet{balona2015}. Note that additional errors have recently been uncovered in the short cadence data processing, which impact both the SAP and PDC data for nearly half of short cadence targets.\footnote{For more information see this erratum:\\ \url{http://keplerscience.arc.nasa.gov/data/documentation/KSCI-19080-001.pdf}}
The amplitude of these calibration errors is typically small, but since the impact for each affected target is not yet known some caution is urged when interpreting the rates of the smallest energy flares. Future versions of this work will utilize Data Release 25 when available in late 2016.

The short and long cadence light curve files were analyzed for every star independently, processing a total of 3,144,487 light curve files from 207,617 unique targets. Since the results from each light curve file are totally independent, this analysis was ideal for parallel computing. To facilitate this large number of light curves I utilized the Western Washington University Computer Science Department's Compute Cluster. This Linux-based cluster has 480 cores, and uses the HTCondor scheduling system \citep{condor-hunter,condor-practice}.

\begin{figure*}[!t]
\centering
\includegraphics[width=3.5in]{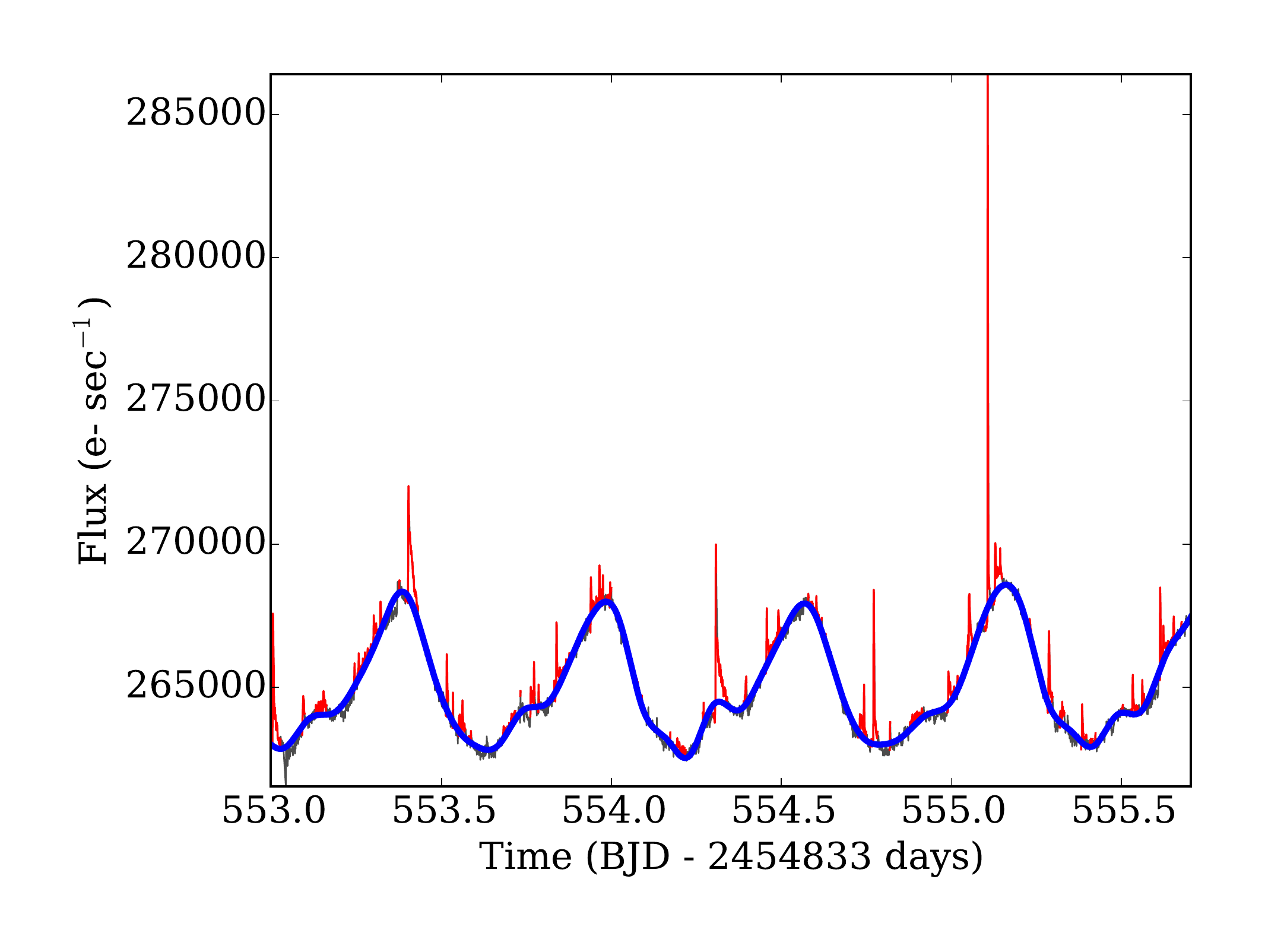}
\includegraphics[width=3.5in]{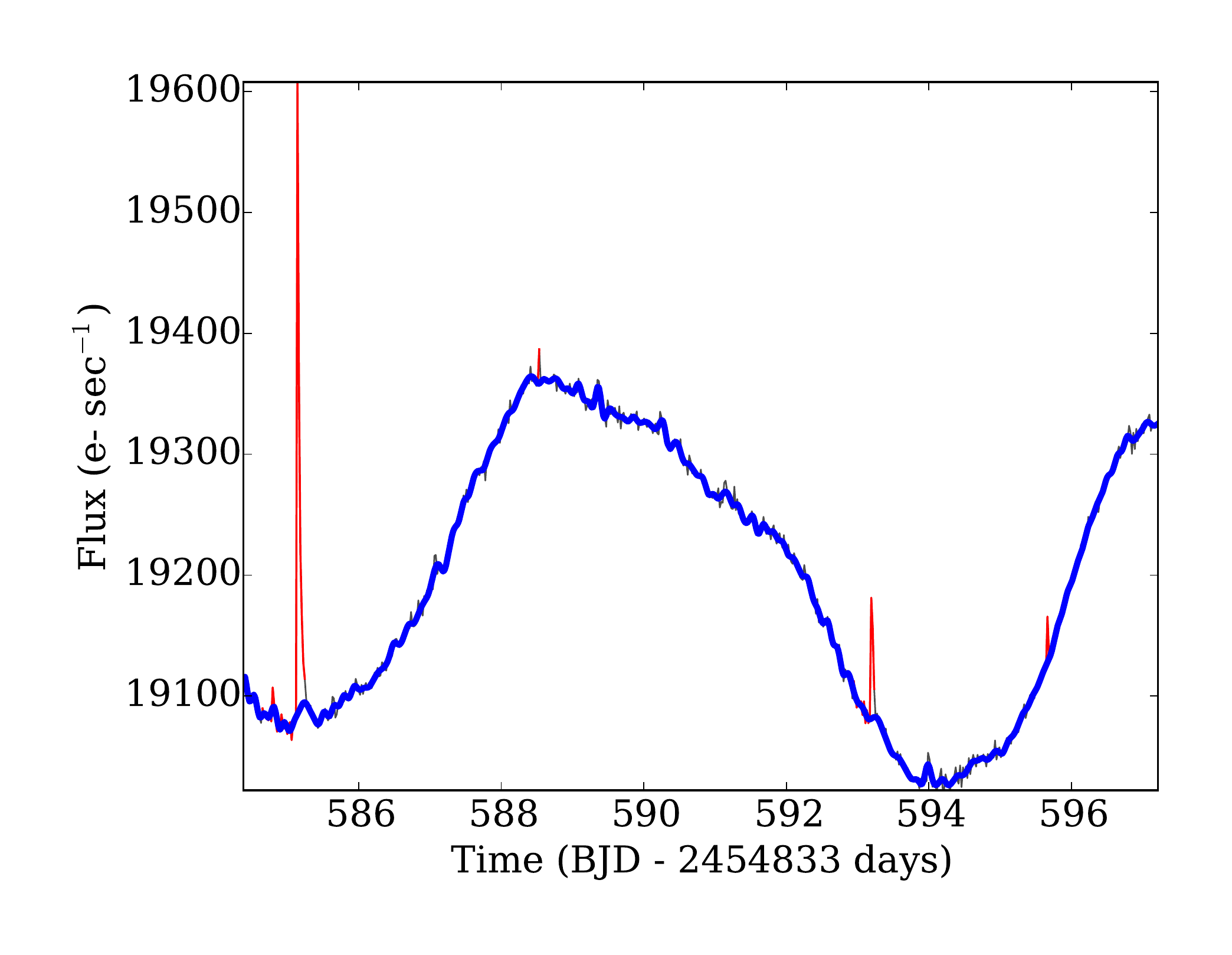}
\caption{
Two examples of flare star light curves we have analyzed. \Kepler SAP\_FLUX is shown (black line) with the final quiescent light curve model overlaid (blue line). Flares recovered in this analysis are highlighted (red lines). Left: Short cadence data from the well-studied M dwarf, KIC 9726699 (GJ 1243). The starspot modulations for this rapidly rotating system is very stable over many rotations. Right: Long cadence data for KIC 6224062. This M dwarf rotates with a moderate period ($\sim$8.5 days), and the starspot configuration evolves significantly in amplitude and phase between subsequent rotations.
}
\label{fig:lc}
\end{figure*}

\section{Flare Finding Procedure}
\label{sec:find}

The process of detecting flares in the \Kepler light curves consists of two steps: 1) building a model for the quiescent stellar brightness over the course of the light curve, and 2) selecting significant outliers from this model as flare event candidates. All light curves from \Kepler contain significant systematic variability due to e.g. spacecraft adjustments and calibration errors. Given the high precision of \Kepler data, astrophysical variability from a variety of physical processes is also observed for many targets on timescales of minutes to days. This combined systematic and astrophysical variability results in a complex variety of light curve morphologies that must be carefully modeled to accurately detect flares. Building this quiescent light curve model for each target, including both long- and short-cadence data, therefore is the most difficult component of this endeavor. The complex, iterative de-trending scheme laid out here has been arrived at from manual experimentation. However, each step in the procedure is designed to remove specific forms of systematic or astrophysical variability.

The entire codebase for this analysis, including all code to generate each figure, is open source and available online.\footnote{\url{http://github.com/jradavenport/appaloosa}}

\subsection{Building the Quiescent Light Curve Model}
\label{sec:find1}
Throughout the description of this procedure each step is numbered for clarity.
{\bf (1)} First, any data points with the SAP\_QUALITY flag bits 5, 8, or 12 set were discarded, which removed epochs with a reaction wheel zero crossing, cosmic ray in aperture, or impulsive outlier detected when co-trending, respectively. The light curve modeling approach begins by subtracting long term variations, which are typically due to systematic errors in the data. Each light curve file, consisting of either an entire quarter of long cadence data or one month of short cadence data, is smoothed via the {\tt rolling\_median} filter from the Python package {\tt pandas} \citep{pandas}, using a kernel size of 1/100$^{\rm th}$ the size of the light curve segment. Additionally, a minimum smoothing kernel size is set at 10 data points, which corresponds to 10 minutes for short-cadence or 5 hours for long cadence data. With this heavily smoothed light curve a 3$^{\rm rd}$ order polynomial is fit, which is then subtracted from the original light curve.

{\bf (2)} Each light curve is then segmented in to regions of continuous observation, breaking the light curve in to individual portions if there are gaps of data of 0.125 days or larger. Each continuous segment was required to be at least 2 days in duration, and any segment less than 2 days in duration was discarded from analysis. These sections are the fundamental regions of data for the analysis because the systematic noise properties of the \Kepler data can change between them due to spacecraft adjustments. As such, the light curve modeling, flare finding, and later the artificial flare injection tests, are all performed on these continuous sections of the light curves.

{\bf (3)} The light curve modeling approach within these continuous segments of data was arrived at through manual experimentation. Within each continuous region the light curve is smoothed using the same {\tt rolling\_median} filter procedure as for the whole light curve, again with a kernel of  1/100$^{\rm th}$ the continuous segment or 10 data points, whichever is larger. This smoothed light curve segment is fit with a 3$^{\rm rd}$ order polynomial, which is again subtracted from the original data.

{\bf (4)} A series of iterative smoothing steps is then preformed to robustly fit the quiescent light curve shape. A 2-pass smoothing with the {\tt rolling\_median} filter with a 2 day kernel is applied, iteratively rejecting flux values residuals that are more than 5 times the \Kepler photometric uncertainty or outside of the 5-95 percentile of the residual distribution.

{\bf (5)} Using this iteratively smoothed light curve segment, which should have most large amplitude flares {\it removed}, I search for periodic signals in the data that are typically due to starspot modulations \citep[e.g.][]{reinhold2013,davenport2015b}.  I use the {\tt LombScargleFast} procedure from \citet{gatspy} to search for periodicity. The largest significant (Lomb-Scargle Power $>$ 0.25) peak in the periodogram is first chosen. If present, the sine function corresponding to this periodic signal is subtracted from the smoothed data. This process is repeated until no significant peak in the periodogram is found, up to a maximum of 5 times.
The search is limited to 20,000 periods spaced logarithmically between 0.1 and 30 days. This multi-period model approach is similar to that used by \citet{reinhold2013a} to search for signals of differential rotation in \Kepler data. The sine curves fit to this data segment are subtracted from the polynomial-smoothed data from step {\bf (3)}, which still has flares present.

{\bf (6)} A 3-pass iterative {\tt rolling\_median} filter approach is then used on the sine-subtracted data, smoothing with a 0.3 day kernel, and iteratively removing outlier points as in step {\bf (4)} above. This again removes the largest energy flares from the light curve.

{\bf (7)} Using this smoothed light curve segment, which should have the starspots mostly removed via the sine-fitting and the flares removed from the median filtering, I perform a 10-pass least squares spline fitting. Rather than removing data points after each pass, the data is iteratively re-weighted \citep[e.g. see][]{green1984} using the per-datum $\chi^2$ statistic multiplied by a penalty factor, $Q$, which is set to a very high value of 400. This results in outliers that increasingly have less and less weight. A similar iterative re-weighting least squares (IRLS) approach was described in the de-trending module of the exoplanet data analysis package, {\tt Bart}.\footnote{\url{http://dan.iel.fm/bart}} Smaller amplitude flares, and the decay phases of larger flares previously removed, are smoothed out at this step.

The final model used to represent the quiescent light curve is defined as the addition of the IRLS smoothed light curve from step {\bf (7)}, and the multi-sine component from step {\bf (5)}. Examples of this model compared with the original data is shown in Figure \ref{fig:lc}.

\subsection{Flare Detection}
\label{sec:find2}

The model generated above is then subtracted from the original data in each continuous light curve segment. I then cross correlated the model-subtracted light curve with a flare profile, using the analytical flare template defined in \citet{davenport2014b}. The flare template is generated with an amplitude arbitrarily set to 1, and a characteristic timescale $t_{1/2}$ of 2 times the local cadence, 60 minutes for long cadence data and 2 minutes for short cadence data. By cross correlating the model-subtracted data with a flare filter we are effectively taking a matched filter approach in detecting flares against the noisy data. Since the cross correlation smooths the flare events out longer in duration, only flare {\it detection} is performed using the matched filter version of the model-subtracted data, and not flare energy measurements.

Candidate epochs belonging to flares are found in this matched filter light curve using a slightly modified version of the FINDflare algorithm, defined by Equations 3a--3d in \citet{chang2015}. This algorithm chooses candidate flares as consecutive epochs are positively offset from the quiescent model by more than the local scatter in the data, as well as being offset by more than the formal errors, where each of these three criteria is governed by scaling factors.
I found that adjusting the scale factor N3, defined in \citet{chang2015} as the number of consecutive points that satisfied the model offset requirements, to N3=2 improved flare recovery for long cadence data and did not negatively impact recovery for short cadence data. The local scatter within each model-subtracted light curve segment in my implementation of FINDflare is determined by computing the median of a rolling 7 data point standard deviation.
To avoid spurious flare detections due to spacecraft reheating, as well as erroneous de-trending, flares are not selected within 0.1 days of the edges of continuous regions of data. Candidate flare events within 3 data points of each other are combined.

Every candidate flare event has several statistics measured and saved for future analysis. These include the start, stop, and peak times of the flare, the maximum amplitude in the original light curve, and the full width at half maximum (in days).
Start and stop times of the flare are defined as the first and last epochs that pass the FINDflare algorithm. This algorithm can under-report the actual flare duration, typically due to the slow decay portion of the flare being mistaken for the quiescent background. While the matched filtering approach mitigates this, the flare durations reported are not exact or based on model fits.
The normalized $\chi^2$ of the flare is then measured, defined as:
\begin{equation}
\chi^2_{fl} = \frac{1}{N}\sum\frac{(y_i - c_i)^2}{\sigma_i^2}
\label{eqn:chisq}
\end{equation}
where $y_i$ is the $i$'th flux value of the flare (using the de-trended fluxes), 
$\sigma_i$ is the $i$'th photometric uncertainty, 
$c_i$ is the $i$'th value from the same region of the iterative quiescent light curve model, 
and $N$ is the number of data points contained in the flare. 
I also compute the 2D Kolmogorov-Smirnov (KS) statistics for the flare, which defines the probability that the flares and some background sample of data are drawn from the same population. The KS test is computed for both the flare data versus an equal sized continuum region around the flare, and the flare versus the de-trended quiescent model. Finally I calculate the flare equivalent duration (ED), which is the integral under the flare in fractional flux units. The ED has units of time (seconds in this case), similar to how equivalent widths of spectral lines have units of wavelength \citep[e.g. see][]{huntwalker2012}. The ED is computed using a trapezoidal sum of the flare data between the start and stop times defined by the FINDflare algorithm.

\subsection{Determining Flare Energies}
\label{sec:find3}

The ED's measured above provide a relative energy for each flare event without having to flux calibrate the \Kepler light curves. As a result the ED's are robust against the observed variability, both systematic and astrophysical. The actual energy of the flare emitted in the \Kepler bandpass (units of ergs) can be determined from the ED (units of seconds) by multiplying by the quiescent luminosity (units of erg s$^{-1}$).

For each star the quiescent luminosity is estimated in order to place the relative flare energies on an absolute scale.
\citet{shibayama2013} accomplish this by assuming blackbody radiation from both the star and flare, as well as a fixed flare temperature of $10,000$ K. However, flare spectra are known to have both non-thermal emission, and changing effective temperatures throughout the event \citep{kowalski2013}. For this reason it better to not assume a single flare spectrum, and instead I estimate the distance and luminosity for each star to determine it's quiescent luminosity.

The Kepler Input Catalog provides ground-based photometry for all available stars in the \Kepler field of view. Using Version 10 of this catalog\footnote{\url{https://archive.stsci.edu/pub/kepler/catalogs/kic.txt.gz}}, I obtained the $g$,  $K_s$, and $Kp$ (\Kepler) photometry for every star in the sample. The $g-K_s$ color is then used to place each star on to a stellar isochrone model, which gives an absolute magnitude and mass for each star. Typical photometric uncertainties from the $g-K_s$ color propagate to mass uncertainties of $\sim$0.02 $M_\odot$.
This assumes that all stars in the sample are on the isochrone's main sequence. A 1-Gyr isochrone from the PARSEC models \citep{bressan2012} with Z=0.019 and no dust extinction is used. Note this will yield an incorrect distance for giant and sub-giant stars. The star's absolute $g, K_s,$ and $Kp$ (\Kepler) magnitudes are determined by linearly interpolating the observed $g-K$ color to the gridded values from the isochrone. The apparent $K_s$ magnitude for each star is used to determine the distance modulus. The isochrone-derived absolute $Kp$ magnitude is finally converted from AB magnitudes to a quiescent luminosity, which is denoted $L_{Kp}$, and is used to convert flare ED's to energies. The resulting flare energy that is calculated does not correct for the spectrum of the flare through the \Kepler bandpass, or for the flare energy emitted outside the \Kepler bandpass, as discussed more in \S\ref{sec:rot}.

\section{Testing Efficiency with Artificial Flare Injections}
\label{sec:fakeflares}

\begin{figure*}[!t]
\centering
\includegraphics[width=3.5in]{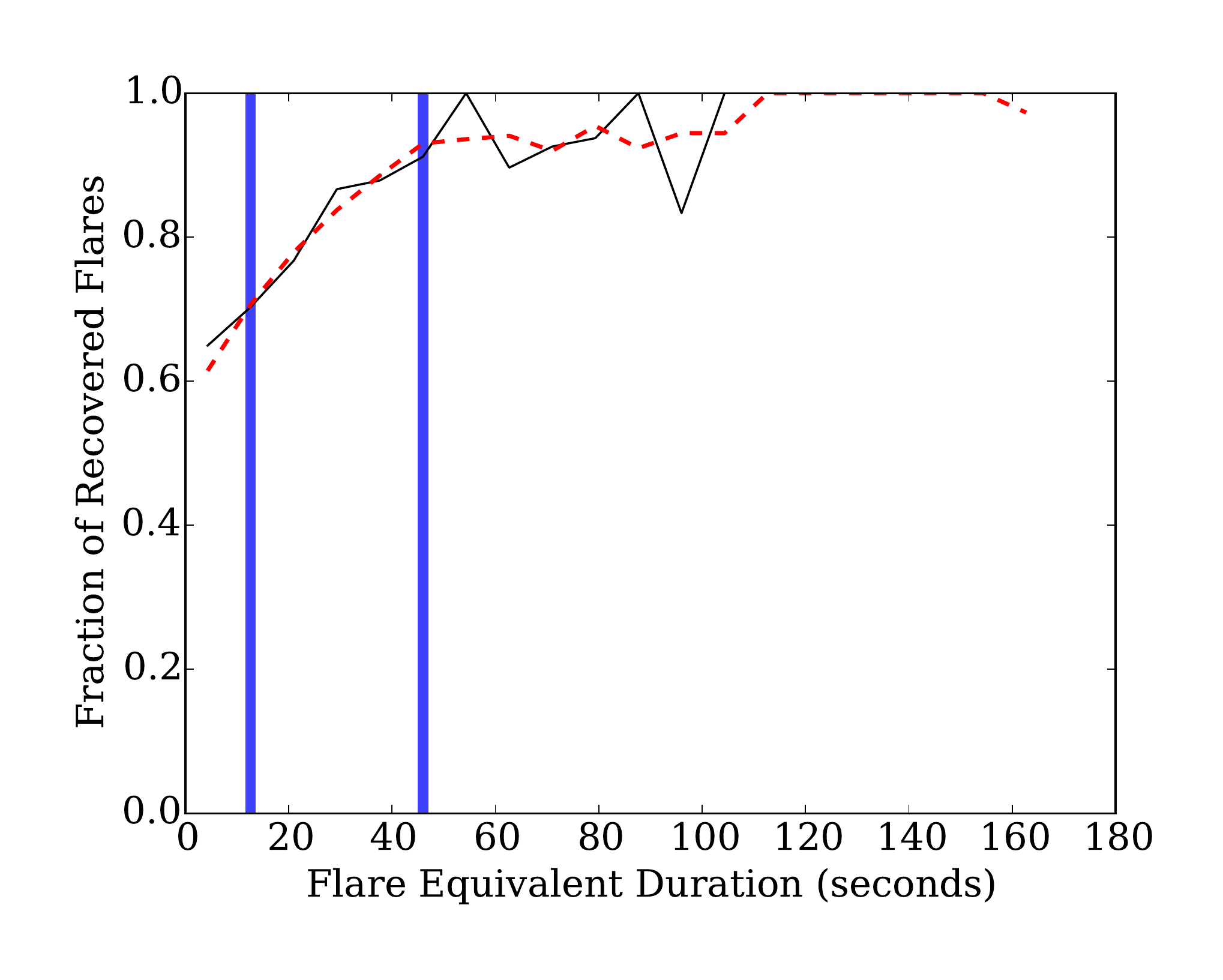}
\includegraphics[width=3.5in]{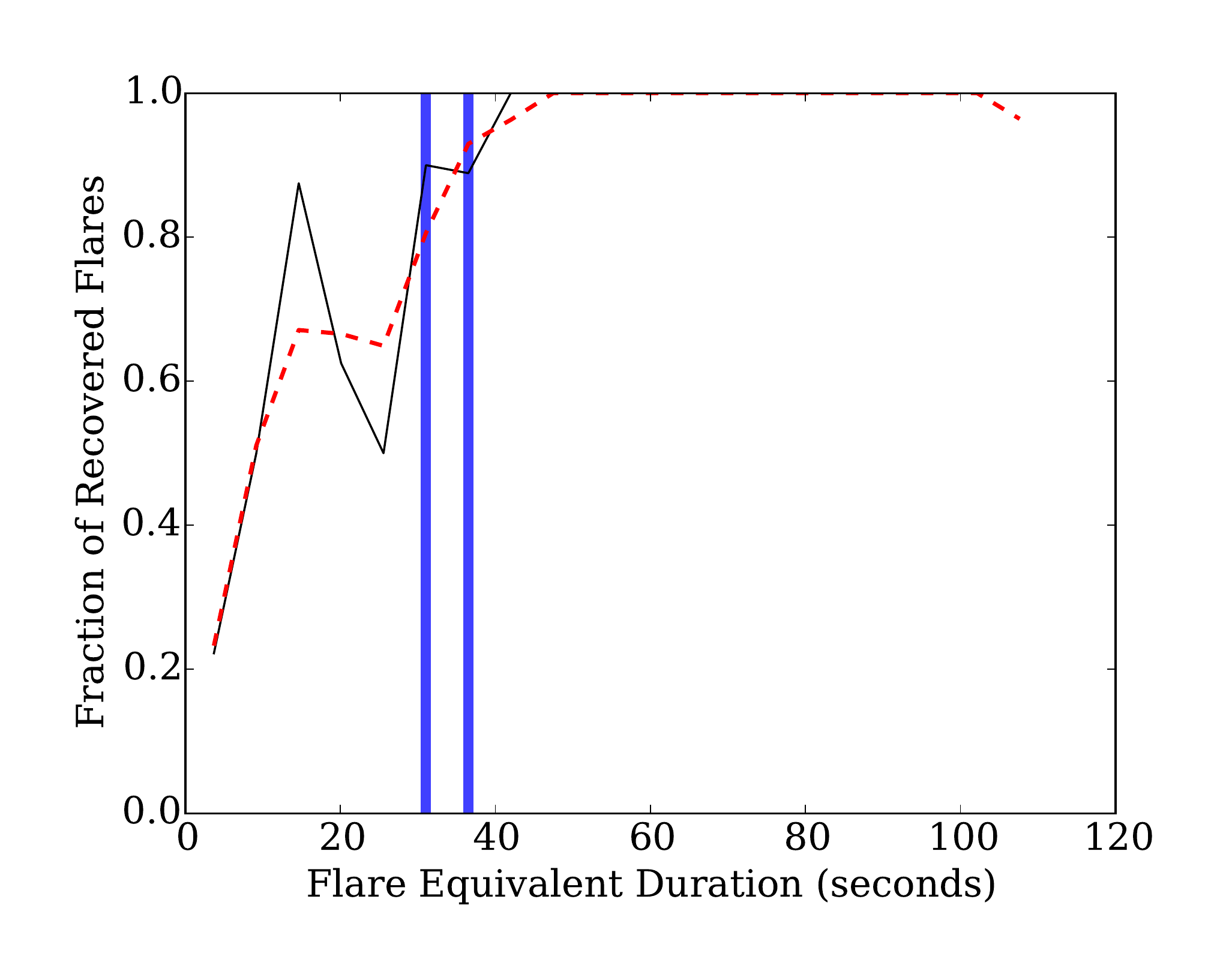}
\caption{
Results from recovery tests of artificial flares injected in to the \Kepler light curves for KIC 9726699 using short cadence (left) and for KIC 6224062 using long cadence (right). The binned recovery fraction for 100 artificial flares is plotted (black line) along with a Weiner-filter smoothed version (red dashed line). Recovery fractions of 68\% and 90\% for the smoothed version are given for reference (heavy blue lines), and are saved for each artificial flare test.
}
\label{fig:fake1}
\end{figure*}

Each continuous section of light curve, defined in Step 2 of \S\ref{sec:find1} above, has unique properties of both systematic noise and astrophysical variability. The accuracy of the de-trending in each light curve section is naturally dependent on the local photometric noise and variability. Comparing flare rates from both long and short cadence data  requires knowing the flare completeness for both cadences, as the sampling rate strongly affects the smallest detectable flares.
Flare recovery efficiency therefore varies between light curve segments, and must be determined within each to accurately characterize the true total flare rate for each star.

Given the variable noise and sampling within each light curve, and the iterative approach of the de-trending procedure used in flare finding, the uncertainty in flare finding cannot be analytically computed. Instead, flare recovery efficiency is empirically determined using artificially injected flares. 
This is analogous to the work of \citet{christiansen2013}, who robustly tested the efficiency in detecting planetary transits from \Kepler data using artificially injected transits. Unlike \citet{christiansen2013} I do not utilize the pixel-level data, and instead inject flares directly in to the raw SAP\_FLUX light curves.

The temporal profile of the artificial flares is the empirical flare model determined in \citet{davenport2014b}. The analytic form of this model (their Equations 1 and 4) describes the flare shape using three free parameters: the impulsive timescale $t_{1/2}$, the flare's peak amplitude, and the time of flare maximum $t_{peak}$.
Within each continuous light curve segment 100 fake flares are injected. The $t_{peak}$ times for the artificial flare events are spaced randomly throughout the quiescent, non-flaring portions of each light curve segment. Each set of 100 fake flares have $t_{1/2}$ timescales chosen randomly in the range $0.5 \le t_{1/2} \le 60$ minutes, and amplitudes between 0.1 and 100 times the median photometric error of the respective light curve segment. While \citet{davenport2014b} show their empirical flare model can be used to identify and decompose complex, multi-peaked flare events, only classical single-peaked events are injected for the artificial flare tests here. The decay phase of the injected flares may partially overlap real or other artificial flares, and as such may create serendipitous complex events.

The light curve segment with added fake flares is then processed using the same iterative de-trending and flare-finding algorithm from \S\ref{sec:find1} and \ref{sec:find2}, respectively. Artificial flares are considered recovered if the flare peak time is contained within the start and stop times of any resulting flare event candidates. For this study I do not keep track of how accurately each artificial flare was recovered, either in duration or event energy. A detailed analysis of the flare energy recovery will be used for evaluating and improving future versions of the code.

The fraction of recovered flares as a function of energy is then computed for each light curve segment. 
Simulated flares are binned as a function of the event energy, using 20 bins of equivalent duration (ED). The locations of these bins were not fixed, and varied between light curve segments due to the simulated flare amplitudes being a function of the local photometric uncertainty. Examples of the recovery fraction for two light curve segments for the M dwarf GJ 1243 (KIC 9726699) are shown in Figure \ref{fig:fake1}. The recovery fraction is then smoothed using a Wiener filter with a kernel of 3 ED bins, and from this smoothed version the 68\% and 90\% flare recovery ED is measured for each light curve segment. These local ED limits are saved along side each recovered flare for use as completeness limits in later analysis. In cases where the 68\% or 90\% recovery rate is not met, a value of -99 is saved for these limits.

\section{The Flare Sample}
In this section I describe the flare sample, including selecting high probability flare event candidates from each light curve, and how to combine both the long and short cadence data to determine robust flare rates.

\subsection{Flare Statistics}

\begin{figure}[!t]
\centering
\includegraphics[width=3.5in]{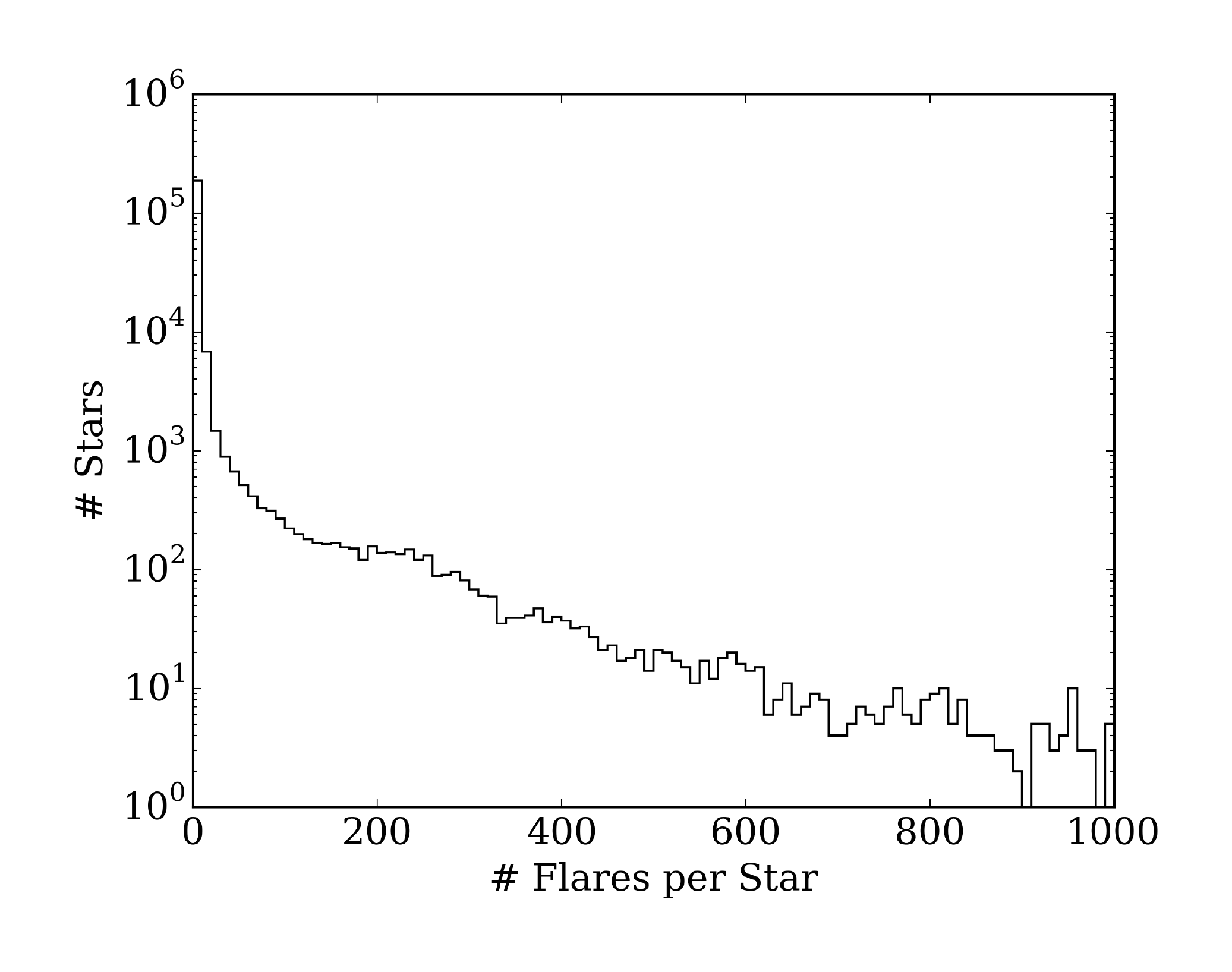}
\caption{
Histogram of number of flare event candidates per star. This includes the entire sample of 2,304,930 flare event candidates from 207,617 stars. For the vast majority of stars the event candidates have small energies and are likely spurious detections that the iterative de-trending algorithm failed to remove.
}
\label{fig:flarehist}
\end{figure}

This analysis of every short- and long-cadence light curve from the \Kepler mission produced 2,304,930 flare event candidates. This large number of events includes event candidates below the 68\% completeness threshold for each light curve segment, spurious detections of non-flares that the iterative de-trending and flare finding algorithm did not remove, and may have detections of real brightening events that are not flares. The distribution of total number of flare event candidates per star is shown in Figure \ref{fig:flarehist}. This distribution reveals that most stars have very few flare event candidates, e.g. only 8149 stars have 25 or more candidate flare events in their light curves. Stars with very few flares are likely to be spurious detections.

Given the large number of light curves and flare events the entire sample could not be manually validated. Instead, further selection criteria was imposed on the sample to analyze only likely flare stars. Specifically each star in the flare rate analysis was required to have:
\begin{itemize}
\item at least 100 total flare event candidates;
\item at least 10 flare event candidates with energies above the local 68\% completeness threshold;
\end{itemize}
Note that these criteria are conservative and will exclude stars with a small number of significant flare detections throughout their light curves.
The Kepler Input Catalog \citep{brown2011a} does provide an estimate of each \Kepler target's surface gravity ($\log g$), which nominally could be used to remove any giants or sub-giants from the sample. However, this $\log g$ estimate has been shown to be unreliable for many stars. I carried out the final analysis including both a cut $\log g \ge 4$ and with no cut on $\log g$. The population statistics explored in the following sections were not strongly affected by this cut, but several known dwarf flare stars from \citet{walkowicz2011} were erroneously tagged as giants and removed. Therefore I opted to not include the $\log g$ cut in the final analysis, but note the sample may include some targets that are not bona fide dwarf stars.

\begin{figure}[!t]
\centering
\includegraphics[width=3.25in]{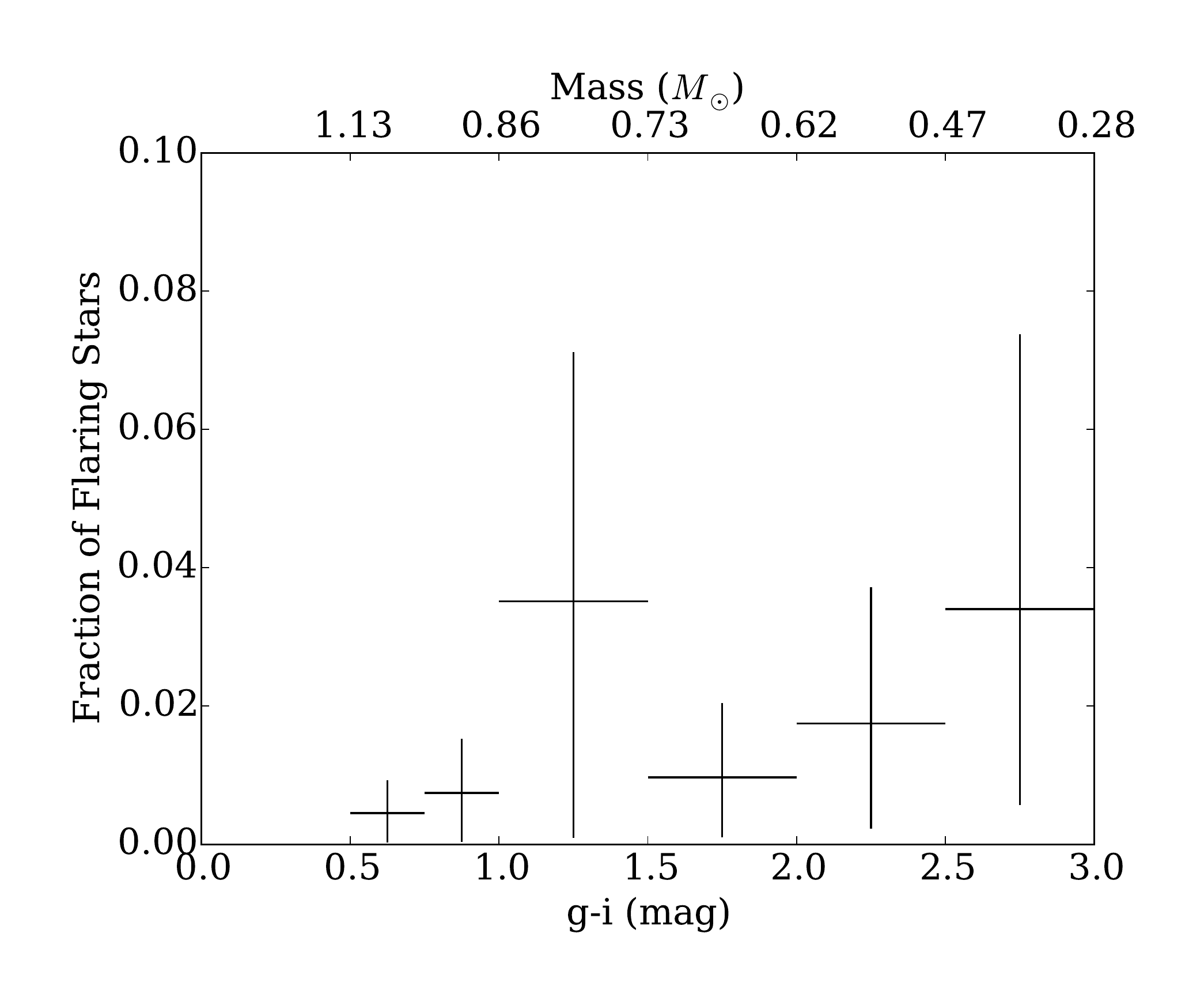}
\caption{
Fraction of stars that pass the final flare sample cuts as a function of their $g-i$ color. Horizontal bars show the range of color within each bin. Vertical uncertainties shown are computed using the 1-$\sigma$ (68\%) binomial confidence interval. A general, but weak trend of increasing total flare occurrence with decreasing stellar temperature (redder $g-i$) is seen.
}
\label{fig:Nvsgi}
\end{figure}

The final sample of flare stars included 4041 targets, or 1.9\% of the stars in the \Kepler data, that passed these selection criteria, with a total of 1,390,796 flare event candidates recovered. From these candidates, 851,168 events (61\%) had energies over the local 68\% recovery threshold determined from the artificial flare injection tests in \S\ref{sec:fakeflares}. Summary statistics for all 4041 stars in this final sample are provided in Table \ref{tbl:datatable}.
Figure \ref{fig:Nvsgi} shows the fraction of \Kepler stars that have detected flares from the final sample of 4041 stars in bins of $g-i$ color, which is a proxy for stellar temperature \citep{covey2007,davenport2014}. The overall fraction of flare stars in the sample (1.9\%) agrees well with total rates from previous studies of \Kepler data, e.g. 1.6\% from \citet{walkowicz2011}. A general trend of increasing rates of flare stars with decreasing stellar mass (redder $g-i$ color) is seen. Flaring M dwarfs, seen in the reddest two color bins make up 2.1\% of the M dwarfs in the \Kepler field. The large, asymmetric uncertainties on flare star occurrence rates in Figure \ref{fig:Nvsgi} are calculated using the 1-$\sigma$ (68\%) confidence interval from the binomial distribution \citep[e.g. see][]{burgasser2003}. These are consistent with the confidence intervals that would be computed from previous \Kepler studies of flare star occurrence rates.

\begin{figure}[!t]
\centering
\includegraphics[width=3.5in]{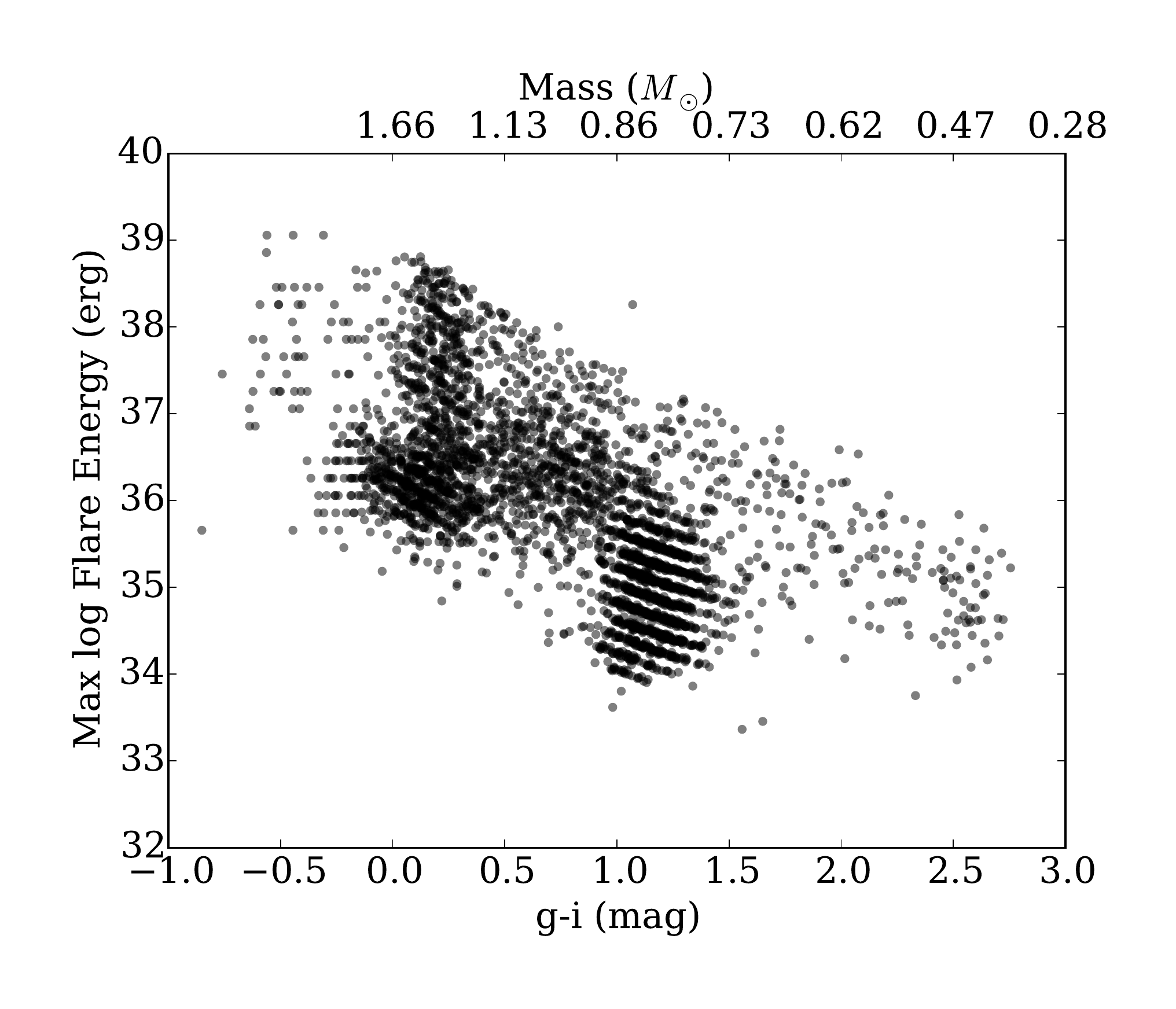}
\caption{
Maximum flare energy per star versus $g-i$ color for the 4041 stars in the final sample. The discretization of flare energies, apparent as ``stripes'' in flare energy in the figure, is due to binning of the flare sample used to combine flare rates between light curve segments.
}
\label{fig:maxcolor}
\end{figure}

The average flare energy detected in the sample is $\log E = 34.6$ erg, very close to the $\sim10^{35}$ erg reported as the average F star flare energy by \citet{balona2012}. Figure \ref{fig:maxcolor} shows the highest energy flare recovered as a function of stellar  $g-i$ color for each star in the final sample. The striping seen is due binning of the flare energy. This binning is used to keep track of flare rates between light curve segments  and for comparing flare energies between stars. 
For comparison, the Sun has a $g-i$ color of $\sim$0.6, and a maximum observed flare energy of $\sim$$10^{32}$ erg \citep{emslie2012}.  Assuming this is the {\it maximum} flare energy the Sun is currently capable of producing, a dearth of objects with similarly low activity levels is recovered in this sample. The sample does contain, however, many G dwarfs that produce super flares.

Most G dwarfs show a peak flare energy of $\sim10^{37}$ erg, consistent with the maximum flare energy found by \citet{wu2015}. However, the highest energy flares in the sample appear to be nearly two orders of magnitude larger than this limit. Note that dust extinction has not been accounted for in the broadband color isochrone-fitting approach for determining the quiescent luminosities. Dust has the effect of making star appear fainter, and thus the distance becomes over-estimated. This may be why larger flare energies are found than in previous studies such as \citet{maehara2015}. The Gaia mission \citep{eyer2013} will provide vastly improved distance estimates for nearly all of these nearby stars, which will help determine the true maximum flare energy observed by \Kepler.

\subsection{Flare Rates from Long and Short Cadence Data}

Flare rates for stars and the Sun have long been described using the cumulative Flare Frequency Distribution \citep[e.g.][]{lme1976}. The ``FFD'' is preferred because flares occur stochastically and span many orders of magnitude in energy and duration. Flare frequency is typically modeled with a power-law function, which shows many small energy flares and very few large energy events. In the the case of the Sun this power-law is traced over $\sim$8 orders of magnitude in observed flare energy \citep{schrijver2012,maehara2015}.

\begin{figure*}[!t]
\centering
\includegraphics[width=3.5in]{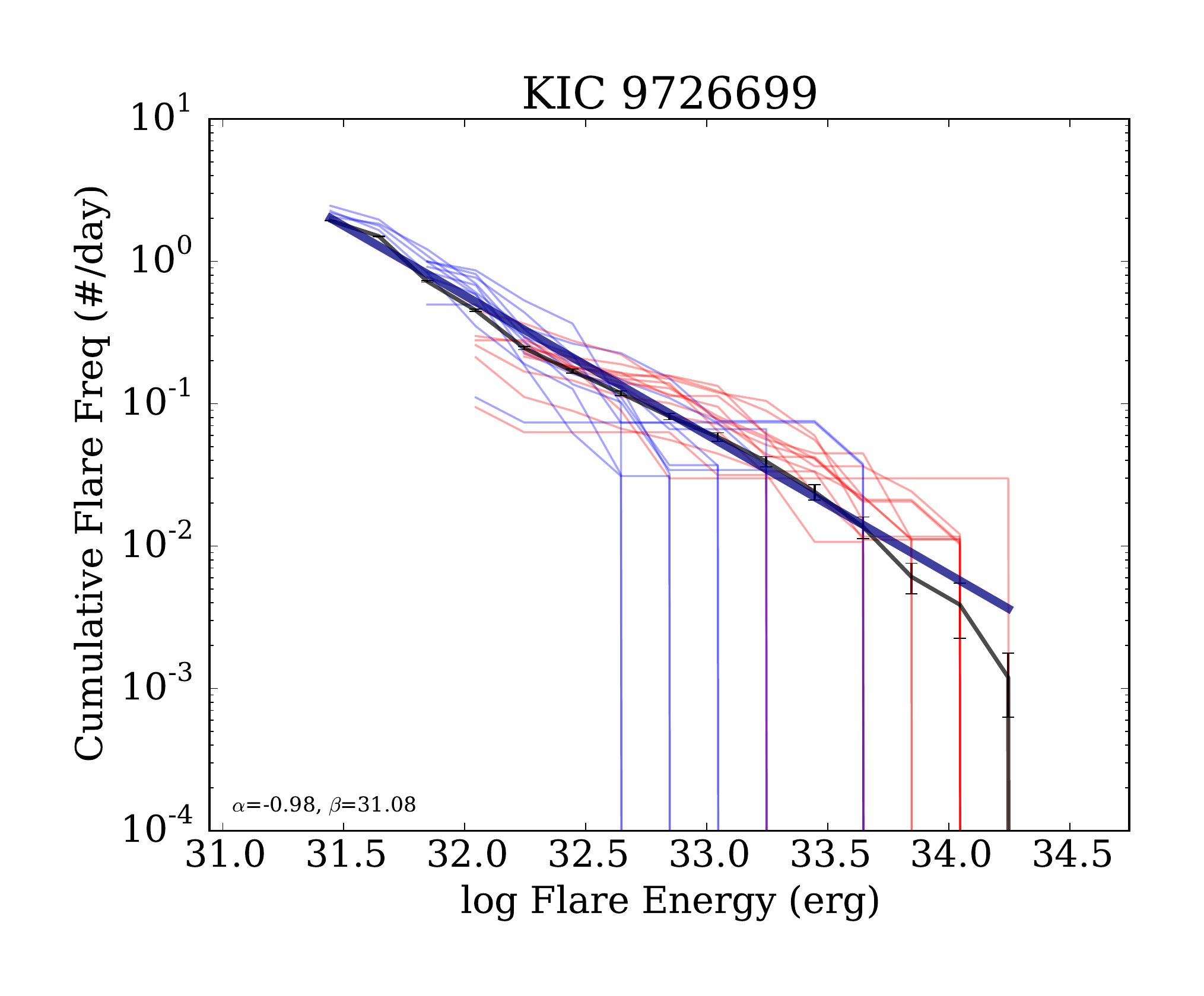}
\includegraphics[width=3.5in]{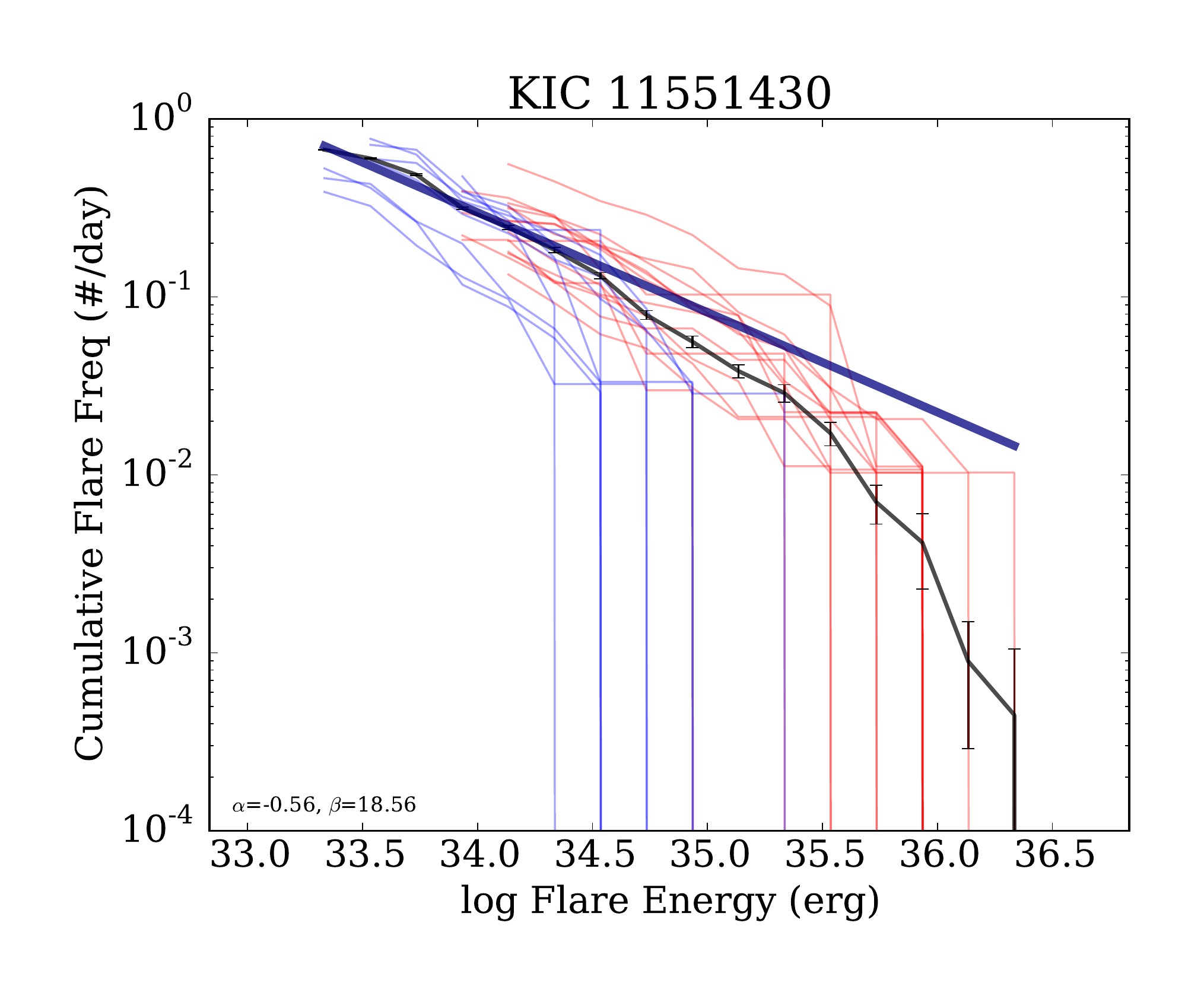}
\caption{
Left: Cumulative flare frequency diagram from all 14 long-cadence quarters (red lines) and 11 short-cadence months (blue lines) for the active M dwarf GJ 1243. The flare rate has been sampled using bins of logarithmic energy. Note the low-energy cutoff for each data file has been set to the average local 68\% flare recovery completeness limit. The average flare frequency distribution is computed by taking the mean in each bin for all files above their respective completeness limits (black line). Uncertainties shown are computed using the Poisson distribution. A weighted least squares power-law fit to the data is computed, which describes well the entire observed flare energy distribution (dark blue line), with power-law fit coefficients listed.
Right: Same diagram for the flaring G dwarf KIC 11551430 (nicknamed ``Pearl'' by David R. Soderblom). 
Unlike GJ 1243, a break is apparent in the flare frequency distribution power-law at high energies.
}
\label{fig:ffd}
\end{figure*}

For all 4041 stars in the final sample a FFD is generated. Unlike \citet{maehara2015} and references therein, I do not produce combined flare frequency distributions for aggregates of stars within spectral type bins, and instead study each star individually. Figure \ref{fig:ffd} shows two examples of FFDs for previously known \Kepler flare stars.

Correctly combining data from different continuous light curve segments (including long- and short-cadence data) to make a single FFD is non-trivial. The varying completeness limits and noise properties mean each light curve segment can potentially probe different flare energy regimes. The artificial flare injection tests allow us to analyze only the range of energies that each light curve segment can detect.
Any effort to search for changes in flare rates over time must take this varying efficiency in to account, or non-physical turnovers or breaks in the FFD may appear.

For each month of short-cadence or quarter of long-cadence data I compute a FFD that is truncated at the low energy end by the average of the local 68\% flare recovery limits defined within that portion of the light curve. These are shown in the two examples in Figure \ref{fig:ffd} color-coded by cadence type. To combine data from these different cadence modes, every FFD is sampled at a fixed set of energies using log-uniform bins of $\log E = 0.1$ erg. The mean flare rate is computed in each FFD bin that has {\it any} valid data (flares above the 68\% completeness threshold). This result in a single FFD for each star, which is overlaid for the two examples in Fig \ref{fig:ffd}. Uncertainties in the flare frequency in this combined FFD are computed for each energy bin using the asymmetric Poisson confidence interval approximations from \citet{gehrels1986}. Each combined FFD is then fit with a weighted least squares power-law, and the coefficients saved for future ensemble analysis.

The FFD for the highly active, rapidly rotating M4 dwarf, GJ 1243, has been previously studied using \Kepler data \citep{ramsay2013,hawley2014,davenport2014b}. These studies have found a constant power law slope describes the FFD up to energies of $10^{33}$ erg using only the short-cadence \Kepler data. In Figure \ref{fig:ffd} I find this power-law extends more than an order of magnitude higher in energy due to the addition of studying the 14 quarters of long cadence data. Unfortunately the iterative flare finding algorithm does not sufficiently recovery flares with energies lower than $\log E \sim 31.5$ erg for GJ 1243. The break in the power-law reported in the human-validated sample from \citet{hawley2014} below $\log E \sim 31$ erg can therefore not verified.

The FFD for the flaring G dwarf, KIC 11551430, shows a remarkable rate of super-flares of nearly 1 per day in the analysis. The highest energy flares for this star are in excess of $10^{36}$ erg. Interestingly, the weighted least squares power-law fit to the FFD for KIC 11551430 in Figure \ref{fig:ffd} shows a significant deviation from a single power-law at the high energy end. Such a break has been suggested for super flare stars previously \citep{chang2015, hudson2015}, and lends weight to the indication by \citet{wu2015} of a maximum flare energy around $10^{37}$ erg for G dwarfs.

\section{Stellar Flares and Rotation}
\label{sec:rot}

Rotation is directly linked to the generation and strength of stellar magnetic fields. Stars lose angular momentum as they age via magnetic braking, which in turn decreases the strength of the stellar magnetic dynamo over time. This age--rotation--activity connection was first illustrated by \citet{skumanich1972}. As a result, the use of rotation periods to infer or constrain stellar ages has recently become popular \citep[e.g.][]{barnes2007,mamajek2008,van-saders2016}.

The decay in magnetic field strength with stellar rotation evolution has also been explored using various magnetic activity indicators. \citet{wright2011}, for example, measured a decrease in the X-ray luminosity as low mass stars spun down, demonstrating a clear connection between the magnetically driven coronal activity and stellar rotation. Similar decay profiles of chromospheric activity with rotation have been observed using indicators such as H$\alpha$ line emission strength \citep{douglas2014}. 

Flares are a highly localized manifestation of stellar surface magnetic fields. The evolution of stellar flare rates and properties with stellar rotation has been explored with limited ground-based flare samples \citep{skumanich1986}. Recent work with \Kepler flares has indicated a decreasing rate of superflares for solar-type stars with increasing rotation periods \citep{maehara2015}. Total flare frequency for \Kepler G, K, and M dwarfs that {\it have} superflares has also been shown to decay with slowing stellar rotation \citep{candelaresi2014}. 
Though a detailed analysis of flare rates with stellar age is beyond scope of this paper, in this section I will point out interesting trends with rotation seen in this sample.

To compare flare rates between stars, the information content within the FFD must be reduced from the two parameters in the power-law fit to a single quantity that describes the star's total flare activity level. Such a metric can be constructed in varying ways. For example, the cumulative rate of flares per day (vertical axis in Fig \ref{fig:ffd}) could be measured at a fixed, standard energy.  While this standardized flare rate metric is not used for the analysis shown here, it is briefly described here for use in future ensemble flare studies. 
Using the average flare energy from the \Kepler sample presented here, a benchmark flare rate could be evaluated at $10^{35}$ erg for all stars. The interpretation of this rate is simple and potentially useful for observers, and its measurement benefits from the careful investigation of flare completeness for each star described in \S\ref{sec:fakeflares}.
However, there are several important limitations in measuring such a quantity. Many stars do not exhibit flares at this particular energy, either for their rarity at such high energies (e.g. flaring M dwarfs), or from faint stars where only the largest super flares are detected. The power-law{\it fit} to the FFD can be evaluated at this benchmark energy, extrapolating the flare rate estimation beyond the observed energy range. However the accuracy of this fit flare rate is limited due to the possible presence of significant breaks in the FFD power-law shape as shown in Figure \ref{fig:ffd} at the high energy end, or by \citet{hawley2014} at lower flare energies. Also, errors in the quiescent luminosity calculation for each star due to factors like interstellar dust correction and isochrone fitting will impact the flare energy estimates, possibly giving inaccurate flare rates at the specified standard energy.

\begin{figure*}[!t]
\centering
\includegraphics[width=2.25in]{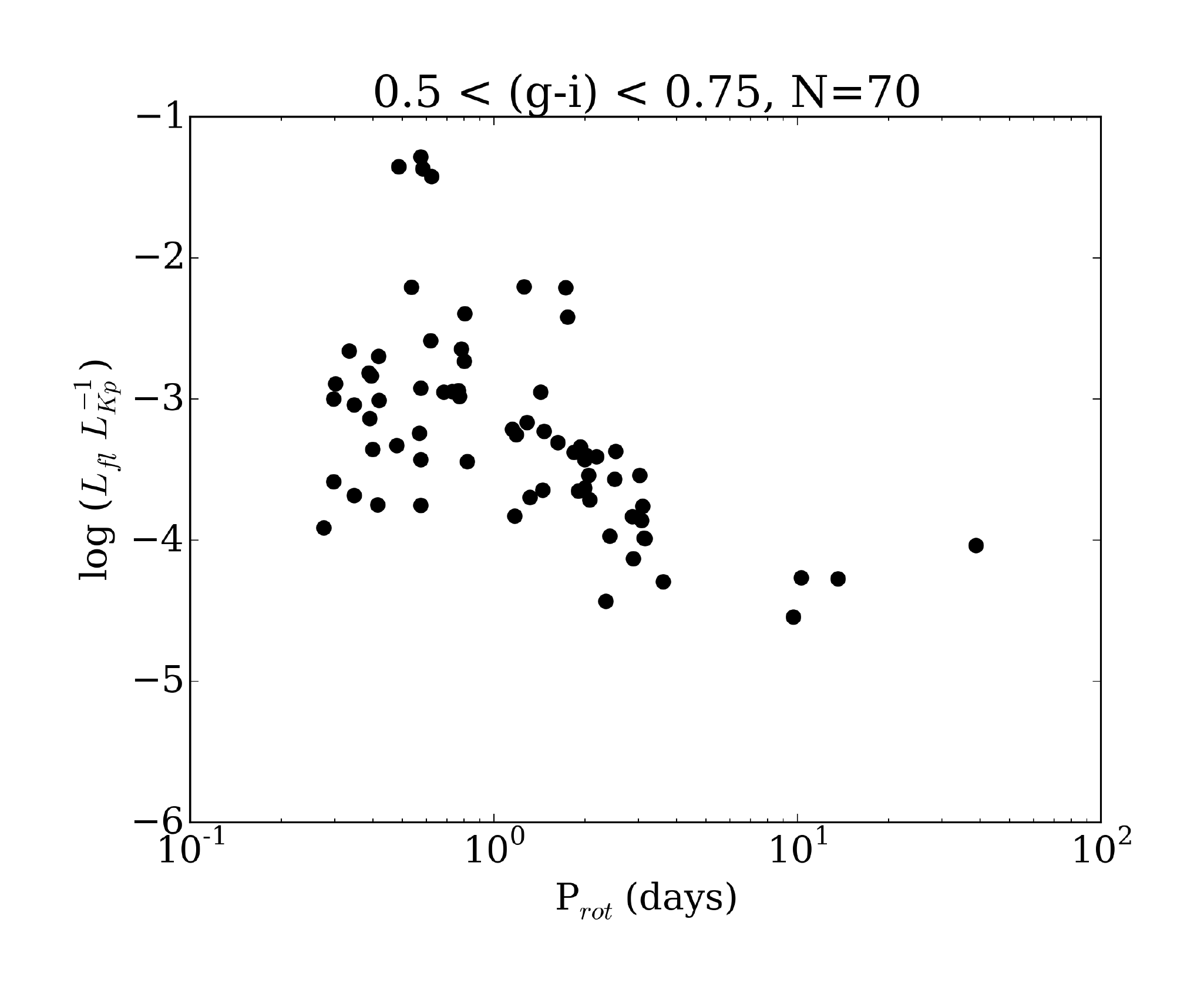}
\includegraphics[width=2.25in]{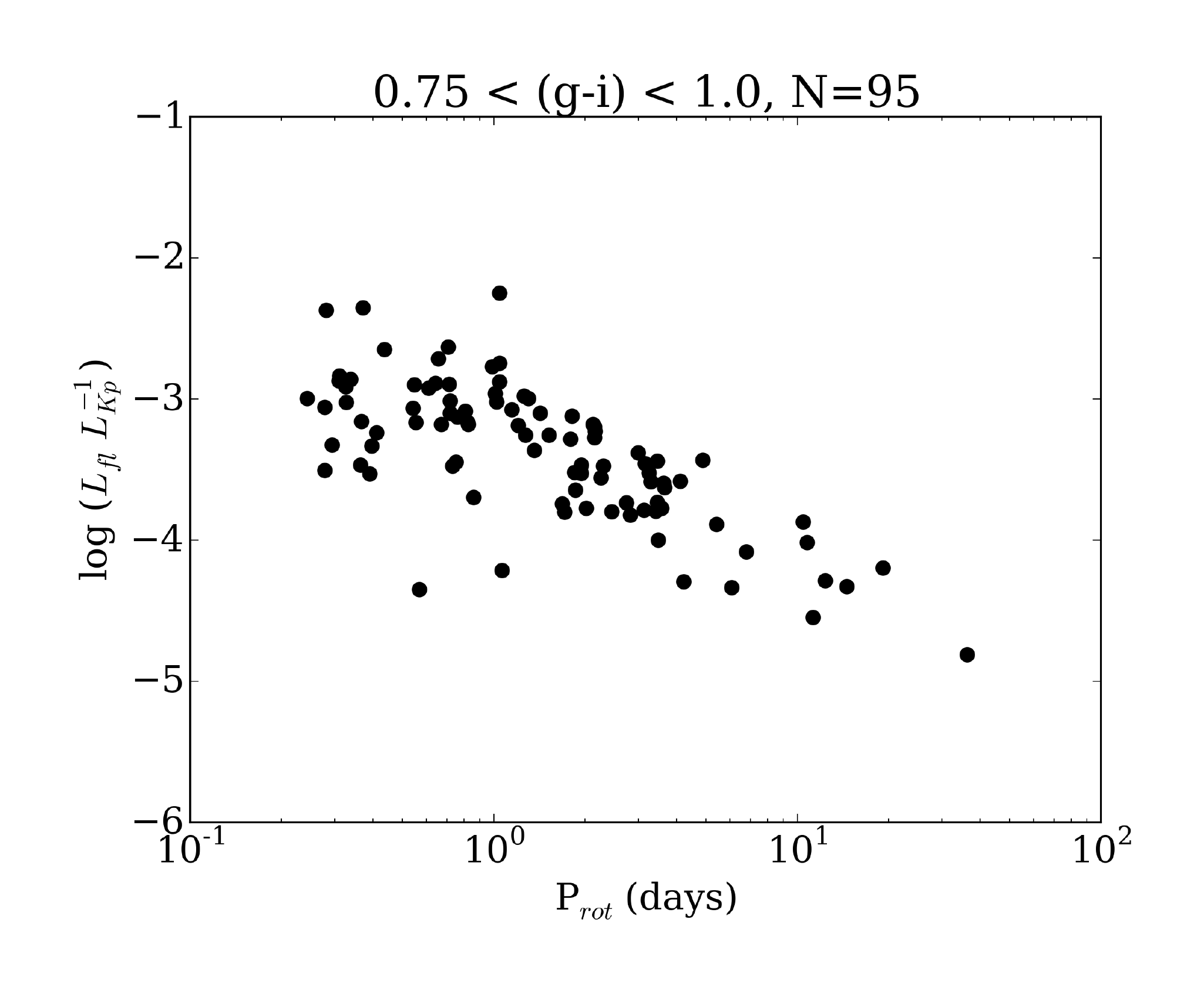}
\includegraphics[width=2.25in]{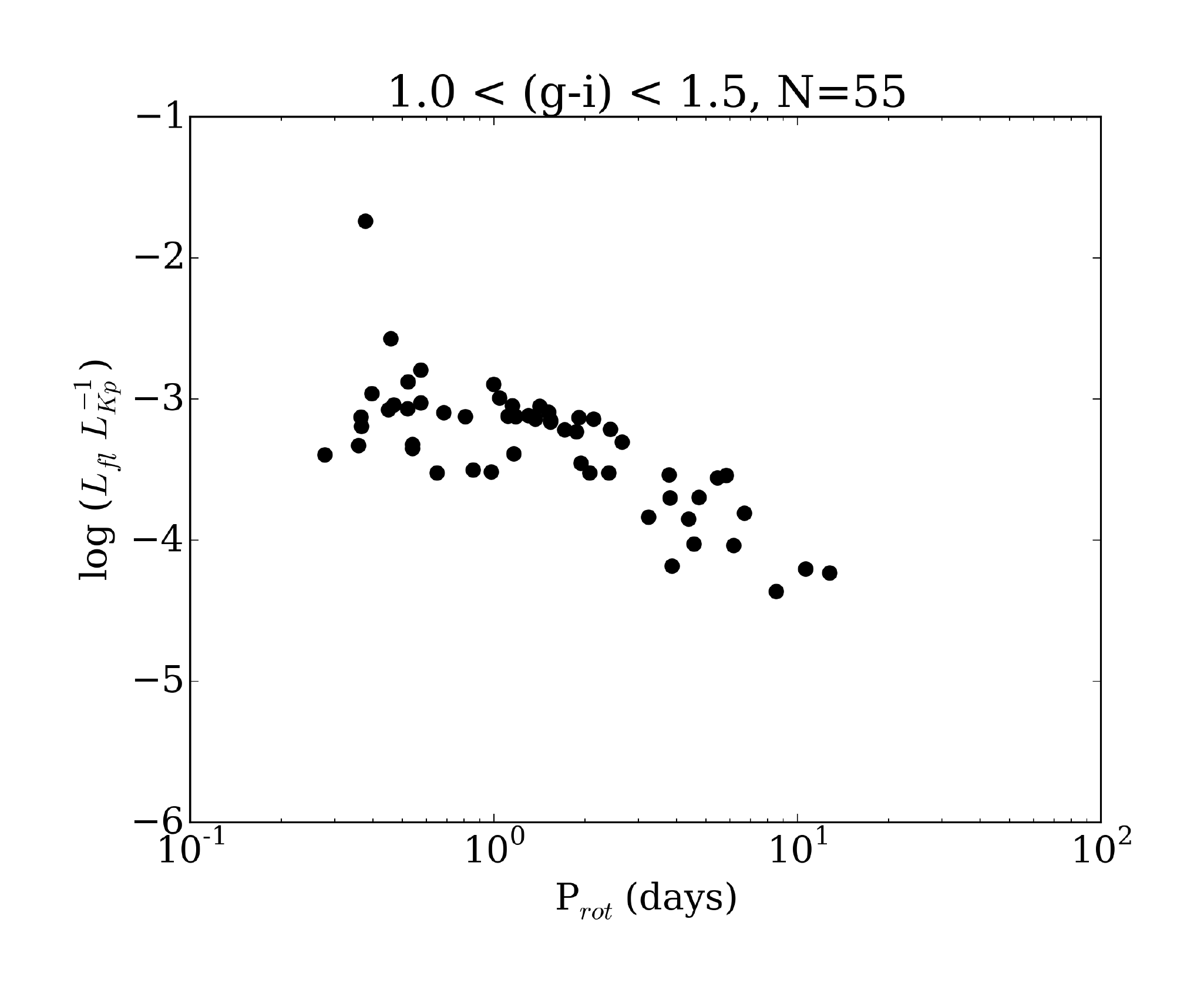}\\
\includegraphics[width=2.25in]{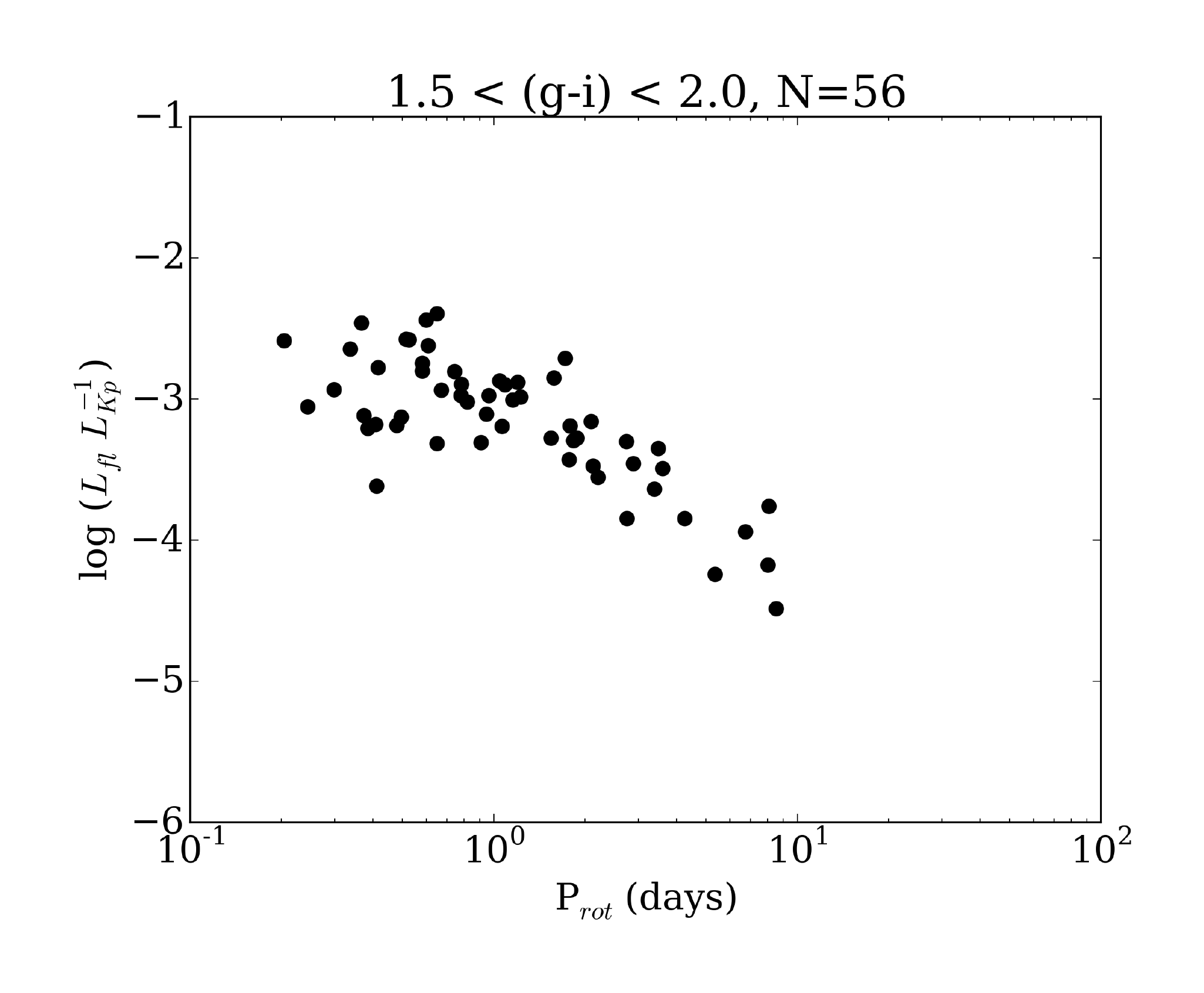}
\includegraphics[width=2.25in]{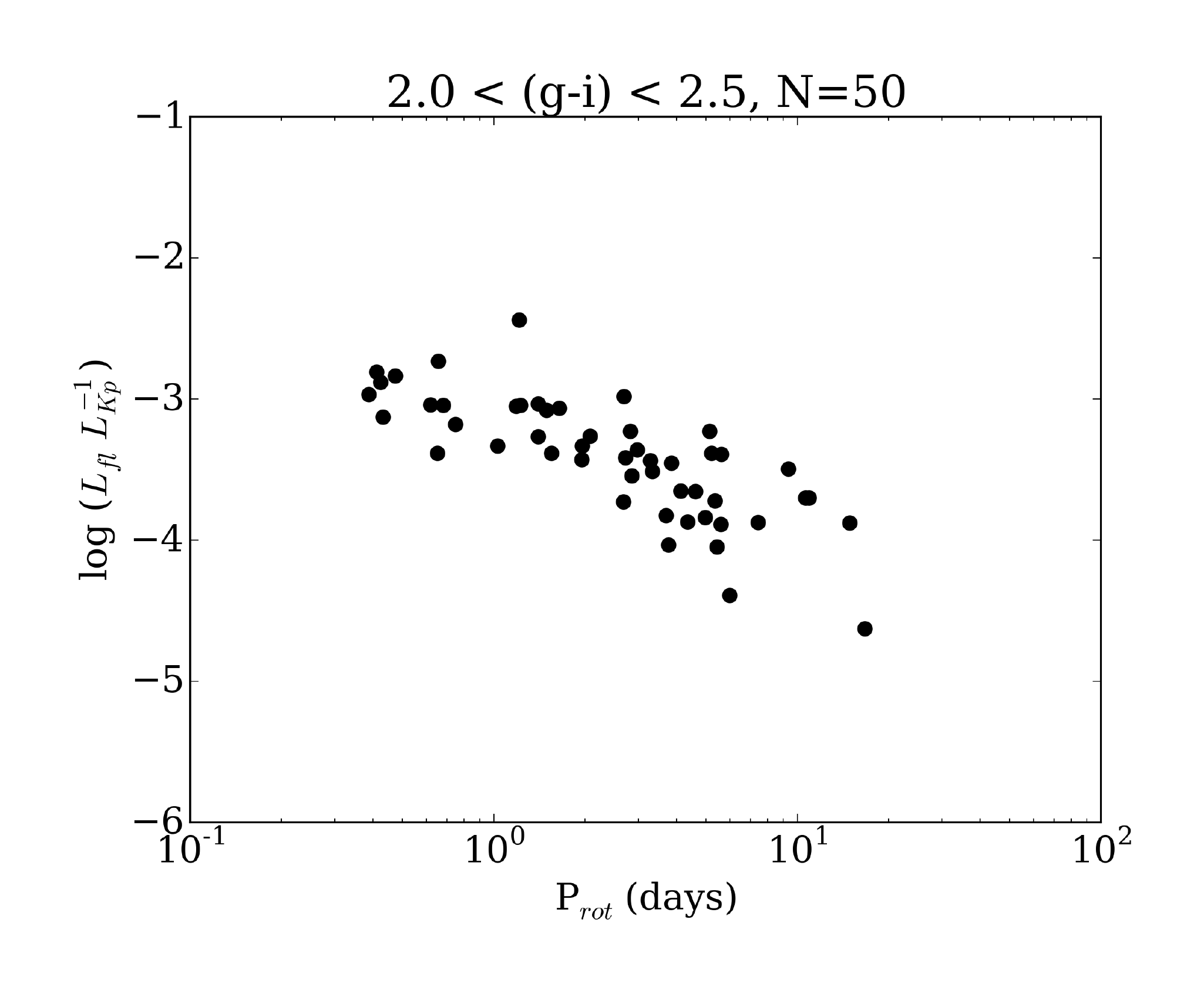}
\includegraphics[width=2.25in]{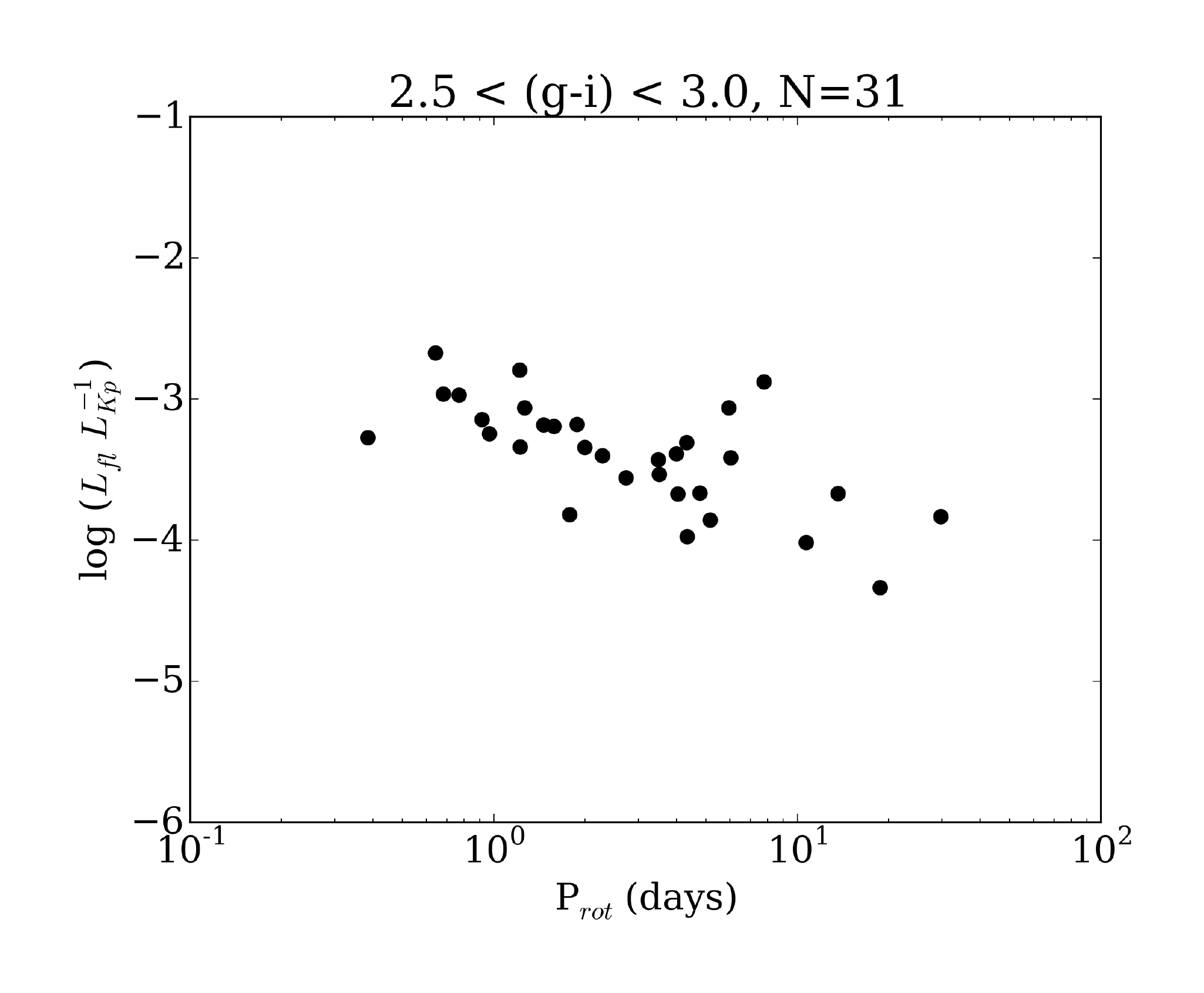}
\caption{
Relative flare luminosity versus rotation period for six cuts in $(g-i)$ color space, which correspond to approximate spectral type ranges of G0--G8, G8--K2, K2--K5, K5--M0, M0--M2, and M2--M4. Each data point represents the total flare luminosity for a star that passes the sample cuts described in the text, and has a valid rotation period from \citet{mcquillan2014}. The number of stars in each bin is indicated in the panel titles. A significant decrease in flare luminosity is seen as a function of rotation period for each subsample.
}
\label{fig:ratecolor}
\end{figure*}

Instead, the total fractional flare luminosity in the \Kepler bandpass, $L_{fl}/L_{Kp}$, is used to characterize each star's flare activity level. This quantity was previously introduced in \citet{lurie2015} to compare the flare yields from the two members of a wide M+M dwarf binary system observed with \Kepler. This metric is calculated by summing up all the flare equivalent durations for each star, and gives the relative luminosity a star produces in flares across the \Kepler bandpass within the observed energy range. This quantity has the advantage of being easily calculated without the need for flux calibrating the light curve or assuming a stellar distance, and is qualitatively similar to other classical indicators of stellar magnetic activity, such as $L_X/L_{bol}$ and $L_{H\alpha}/L_{bol}$.

Note that this quantity could be normalized to the stellar bolometric luminosity, by computing $L_{Kp}/L_{bol}$. Generating this normalization would be analogous to the creation of the ``$\chi$ factor'' used to convert H$\alpha$ equivalent widths in to $L_{H\alpha}/L_{bol}$. The H$\alpha$ $\chi$ factor accounts for the changes in the spectral continuum shape and contrast between stars of different spectral types. A comparable ``flare $\chi$'' to convert $L_{fl}/L_{Kp}$ in to $L_{fl}/L_{bol}$ would require both a correction for the stellar spectrum across the \Kepler bandpass, as well as an estimation of the spectral energy distribution of the flare throughout the event. This latter term requires a unified model of white light emission for both simple (single-peaked) and complex (multi-peaked) stellar flares \citep{allred2015}.

The uncertainty for $L_{fl}/L_{Kp}$ is calculated by adding in quadrature the uncertainties on the equivalent duration from every flare. This uncertainty on equivalent duration for each flare is computed as:
\begin{equation}
\sigma_{\rm{ED},i} =  \sqrt{\frac{\rm{ED}_i^2}{ n_i \chi_i^2}}
\end{equation}
where ED$_i$ is the flare's equivalent duration, and $n_i$ the number of data points contained in the flare. The $\chi^2$ here is the typical reduced goodness-of-fit metric computed for each flare in Equation \ref{eqn:chisq}. In this way, which may be counter-intuitive, larger values of $\chi^2$ indicate {\it more} certainty in flare detection, and in turn yield a smaller error on the total $L_{fl}/L_{Kp}$ computed for a star.

From the final sample of 4041 flare stars, 402 targets had rotation periods of at least 0.1 days measured from the ensemble analysis of \citet{mcquillan2014}.
These rotation periods were determined using the auto-correlation function, which is less prone to detecting period aliases as compared to Lomb-Scargle approaches. These periods have been well vetted, and compared against independent measures of rotation in the \Kepler data \citep{reinhold2013}. Additionally, the sample of stars with reported rotation periods from \citet{mcquillan2014} are not known to have significant contamination from giant stars. 
While the de-trending and flare detection algorithm featured in this work (\S\ref{sec:find1}) does fit sine curves to the the continuous portions of the light curve, at present it does not report a characteristic period for each object. Future work with updated version of the algorithm and newer releases of \Kepler data will investigate the possible correlation between the \citet{mcquillan2014} rotation periods and the periods determined by this de-trending algorithm.

In Figure \ref{fig:ratecolor} I show the relative flare luminosity versus rotation period for the 402 stars with valid periods, separated in to six bins of the stellar $g-i$ color. Using Table 4 from \citet{covey2007}, these $g-i$ color bins correspond to spectral type ranges of G0--G8, G8--K2, K2--K5, K5--M0, M0--M2, and M2--M4, respectively. In total 357 stars fall within the color bins shown in Figure \ref{fig:ratecolor}. There were 45 additional objects with KIC colors bluer than $g-i=0.5$, i.e. with spectral types of A and F. While it is surprising to detect flares or flare-like events from such early type stars given their lack of deep convection zones, they have been reported previously in the \Kepler data \citep{balona2012}.

The earliest spectral type (bluest) bin in Figure \ref{fig:ratecolor} shows only a weak correlation between relative flare luminosity and stellar rotation period. The large scatter in this diagram, especially for the stars with very high levels of flare activity, may be due to outliers in the sample from binary stars, or stars with anomalous flare-like events as seen in the A and F stars noted above. However, stars in this mass range with rotation periods less than $\sim$10 days are also considered to be in the ``super-saturated'' dynamo regime \citet[e.g.][]{argiroffi2016}. Stars with saturated dynamos have a high level of magnetic activity, and show a decoupling between magnetic activity indicators and their rotation periods. The mechanism behind  the observed magnetic activity saturation is debated. Given the lack of G dwarfs with long rotation periods in \citet{mcquillan2014}, and thus in the sample of 402 stars presented here, it is not clear that any strong or coherent evolution in flare activity with rotation should be expected for this bluest bin.

\begin{figure*}[!t]
\centering
\includegraphics[width=5in]{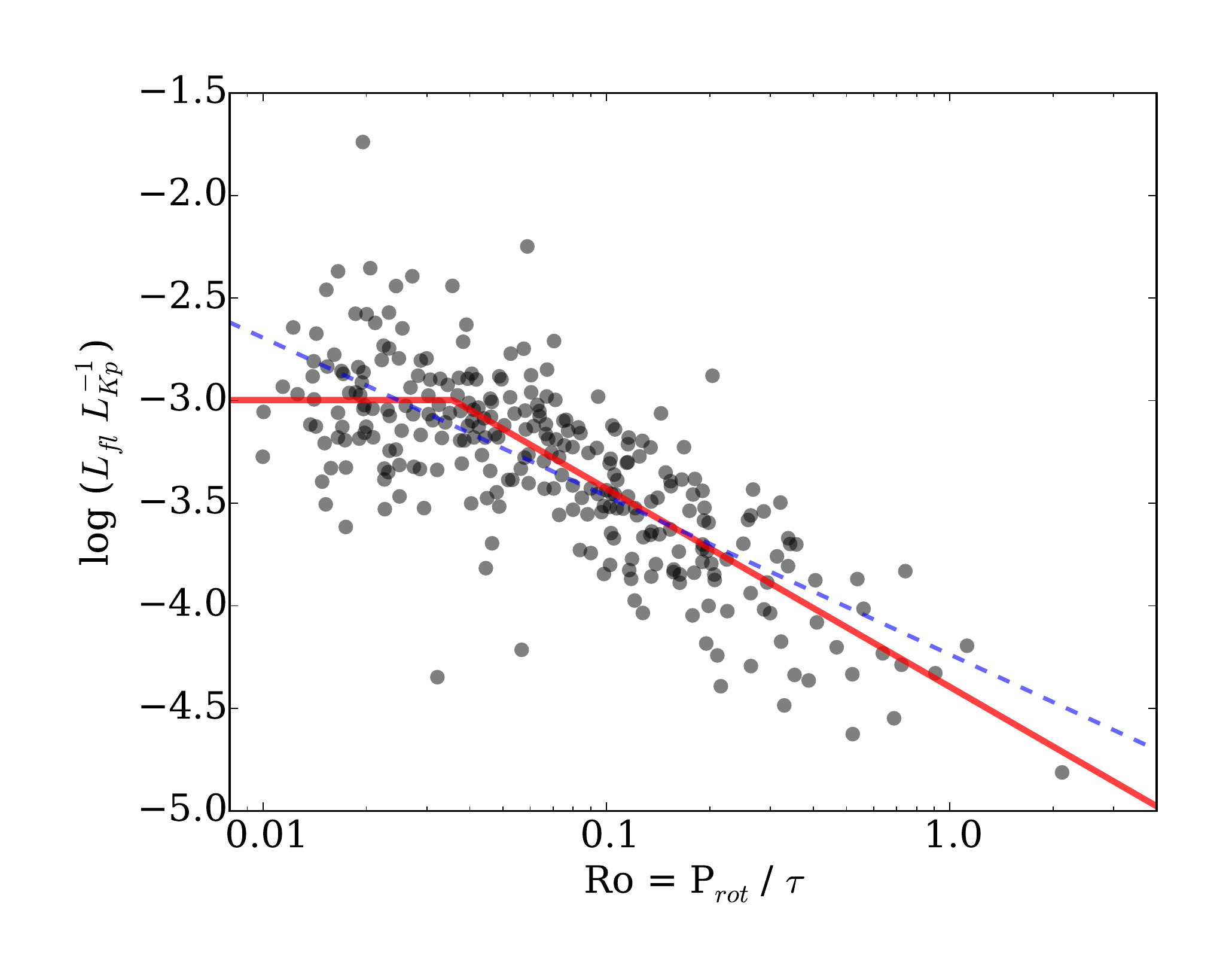}
\caption{Relative flare luminosity versus Rossby number (Ro) for the final sample of flare stars in the color range $0.75 < g-i < 3$. Convective turnover timescales ($\tau$) are derived from Eqn. 11 of \citet{wright2011}. Uncertainties in the total relative flare luminosities, described in the text, are smaller than the data points shown. A clear trend is seen in this diagram, with flare activity decreasing at larger Rossby numbers. Two models are shown for comparison: a single power law with slope of -0.77 (blue dashed line) and a broken power law (red solid line) as is typically used to describe magnetic activity versus Rossby number. The ``saturated'' regime suggested by the latter model occurs at Ro$\sim$0.03, and a power law decay with slope $\sim$-1 dominates to high Ro.
}
\label{fig:rossby}
\end{figure*}

For stars with $g-i>0.75$ (spectral types later than approximately G8) in Figure \ref{fig:ratecolor}, a significant trend in flare activity is seen with rotation period. A saturation-like regime is seen at short periods, and power law decay for rotation periods longer than $\sim$1 day. For stars in the reddest bin ($g-i>2.5$, spectral type M2--M4), the paucity of targets with very short rotation periods means only a power law decay is observed. There are too few stars with spectral types later than M4 to investigate the evolution of flare activity with rotation across the ``fully convective boundary''. This form of saturation and decay profile of magnetic activity has been observed using several other metrics. X-ray luminosity for low-mass stars  saturates at rotation periods of a few days \citep{pizzolato2003,wright2011}. Ultraviolet excess emission appears to follow X-ray luminosity for young stars, with a similar saturation regime \citep{shkolnik2014}.

Stellar activity indicators are often compared between low-mass stars with a range of masses by normalizing the rotation period to a dimensionless rotation indicator. The Rossby number is commonly used for this purpose, and is defined as Ro=P$_{rot}/\tau$, where $\tau$ is the (model derived) convective turnover timescale that is a function of stellar mass.  In this way Rossby number gives a mass-independent metric for the star's rotation, which is useful for comparing to manifestations of magnetic activity. For example, \citet{candelaresi2014} have investigated superflare rates in \Kepler as a function of Rossby number. Masses for stars in the final flare sample presented here are determined using the isochrone fits described in \S\ref{sec:find3}. The $\tau$ values are computed using Eqn. 11 from \citet{wright2011}, which are then used to convert rotation periods from \citet{mcquillan2014} in to Rossby number.

In Figure \ref{fig:rossby} I present the relative flare luminosity as a function of Rossby number for stars with spectral types later than G8. A clear decay in flare activity with increasing Rossby number (or rotation period) is seen. Following other studies of activity evolution with Rossby number \citet[e.g.][]{wright2011}, a simple piecewise model can be used to fit the data in Figure \ref{fig:rossby}, with a constant (flat) level of activity up to a critical Rossby number, and a single power law decay for larger values of Ro. The data in Figure \ref{fig:rossby}were fit using this piecewise function and a weighted least squares fitting routine, yielding saturated relative flare luminosity, critical Rossby number, and power law slope values of:
\begin{eqnarray}
[\log (L_{fl} L_{Kp}^{-1})]_{\rm sat} =& -2.99 \pm 0.03 \notag \\
\rm{Ro_{sat}} =& 0.036 \pm 0.004 \notag \\
\beta = & -0.97 \pm 0.06 
\end{eqnarray}

\noindent
respectively. The critical Rossby number separating the saturated and decay regimes of Ro$_{\rm sat}$ is much smaller than the typical value of 0.1 found using X-ray activity, indicating stellar flares become coupled to a star's angular momentum evolution {\it sooner} than the coronal X-ray emission \citep{pizzolato2003}. \citet{wright2011} point out that the saturation threshold Rossby number is not universal among chromospheric and coronal activity indicators, and that \citet{marsden2009} find a break as low as Ro$\sim$0.08 using Ca II emission.

The power law decay in flare luminosity shown in Figure \ref{fig:rossby} is slower than for X-ray luminosity or $L_{X}/L_{bol}$, which typically is found to decay with a power law slope of $\beta \sim-2$ \citet{wright2011}. A similarly shallow decay with Rossby number of  $\beta \sim-1$ was indicated for chromospheric H$\alpha$ emission in two open clusters by \citet{douglas2014}. Flare activity has been suspected as a cause for the heating of both the stellar chromosphere and coronae \citep{skumanich1985}, and flares have repeatedly been shown to be a probable cause of quiescent coronal emission \citep[e.g.][]{kashyap2002}. The similar evolution of H$\alpha$ emission and flare activity found in this work is further suggestion towards a connection between flares and chromospheric heating.

The data in Figure \ref{fig:rossby} can also be fit using a single power law decay, with no saturation regime. Using this model a power law decay slope of $\beta = -0.77 \pm 0.04$ is found. This single power law has nearly the same quality of fit as broken power law model using the reduced $\chi^2$ parameter. The Bayesian Information Criterion (BIC) can be used to determine which model is preferred by penalizing additional degrees of freedom or parameters in the model. A more complicated model is typically preferred if the BIC improves by at least 2. I calculated the BIC for both the single and broken power law models as: BIC = $\chi^2 + k \times \ln(n)$, where $k$ is the number of free parameters in the model and $n$ the number of data points contained in Figure \ref{fig:rossby}. The broken power law model had a BIC value 6\% {\it larger} than the single power law, indicating the simpler model is slightly preferred for this data.  

Interestingly, when each sub-sample shown in Figure \ref{fig:ratecolor} is fit with these two models, the picture becomes less clear. The broken power law model is preferred by from the BIC for the two bluest (highest mass) samples, while the single power law model is slightly preferred for the reddest two (lowest mass) samples. As the statistical errors on $L_{fl}/L_{Kp}$ are far smaller than the scatter shown in Figures \ref{fig:ratecolor} or \ref{fig:rossby}, it is not clear if the change in flare activity with Ro can be described by either the single or broken power law model for all stars.

\section{Summary and Discussion}

I have presented a homogeneous search for stellar flares using every available light curve from the primary 4-year \Kepler mission. A final sample of 4041 flare stars was recovered, with 851168 flare events having energies above the locally determined completeness limit.
This analysis included extensive completeness testing, using artificial flare injection and recovery tests throughout each light curve to determine the flare recovery efficiency as a function of time.  
While these tests provide a robust and straightforward means to estimate the event recovery efficiency, they currently do not estimate how accurately artificial flare event energies were reproduced. Future improvements to the flare finding algorithm could keep track of the recovered energy and duration for every simulated flare. The light curve de-trending algorithm may also be simplified by using more advanced techniques, such as continuous autoregressive moving average-type models to describe the many forms and timescales of variability at once \citep[e.g.][]{kelly2014}.

As a demonstration, in Figure \ref{fig:ffd} I have shown one example of a deviation or break from a single power law in flare occurrence at large flare energies. However, many other active stars show similar breaks at large flare energies in this sample. A systematic follow-up study of FFDs is needed to determine if this break is common among young Solar-type or low-mass stars, which will be impact detailed studies of superflare occurrence. 
The maximum flare energies recovered in this work are also much higher than previous studies, with a small number of stars in Figure \ref{fig:maxcolor} exhibiting up to $10^{39}$ erg events. These events may be the result of errors in either the light curve de-trending leading to spurious flare events, or the quiescent luminosity determination yielding incorrect energies for real events.
 Note also that small offsets between flare energies calculated with short- and long-cadence data are seen, as in Figure \ref{fig:ffd}. This may be largely an effect of the respective light curve sampling \citep[e.g. see][]{maehara2015}.

From the final sample of 4041 flare stars, 402 were found to have published rotation periods from \citet{mcquillan2014}.
 A striking evolution of flare activity with stellar Rossby number is seen. This evolution includes a possible saturated flare regime for rapidly rotating (low Rossby number) stars, and power-law decay that is qualitatively similar to previous results for chromospheric H$\alpha$ emission. The tentative discovery of a flare saturation regime gives credence to the model of magnetic activity reaching a peak level due to a maximum filling factor of small scale active regions on the surface \citep{vilhu1984}. However, the Rossby saturation limit (Ro$_{sat}$) and the power-law decay slope do not match expected values from most previous studies of magnetic activity saturation and evolution. Since the sample of flare stars is biased more towards K and M dwarfs than most studies of coronal or chromospheric saturation, the smaller Ro$_{sat}$ value may indicate lower mass stars have different saturation limits than Solar-type stars \citep{west2009}. Alternatively, this result may indicate flare activity traces a fundamentally different component of the stellar surface magnetic field. The connection between white light flares, chromospheric emission, coronal heating, and the generation of the magnetic dynamo clearly deserves further observational investigation. Given the varied dependance on Rossby number that these related manifestations of magnetic activity have shown, the dependence of Rossby number as the fundamental metric for tracing dynamo evolution is uncertain \citep{basri1986,stepien1994}.

The large sample of flares observed by \Kepler enables a new generation of statistical studies of magnetic activity. This may yield power advances in constraining stellar ages via flare rates or maximum flare energies, known as ``magnetochronology''. The uniformity of flare activity evolution can be tested using wide binary stars or stellar clusters, many of which are being observed by the \Kepler and K2 missions. 
Beyond the total flare activity levels for ensembles of stars, the temporal morphology of individual flare events may shed new light on the formation of ``classical'' versus ``complex'', multi-peaked flares, as discussed by \citet{davenport2014b}, \citet{balona2015}, and \citet{davenport2015c}. Modeling the detailed structure of these complex events will help in detecting rare ``quasi-periodic pulsations'' in flares \citep{pugh2015}.
Finally, the statistical knowledge we gain from \Kepler will enable more accurate predictions of flare yields from future photometric surveys.

\section*{Acknowledgements}

I wish to thank Austin C. Boeck, Riley W. Clarke, and Jonathan Cornet for their help in manually inspecting several flare light curves in the creation of this codebase. Suzanne L. Hawley and Kevin R. Covey provided invaluable discussions of magnetic activity, and feedback that greatly improved this work. I thank D. Foreman-Mackey and D. Hogg for making their various light curve de-trending algorithms publicly available, and for their extensive discussions and resources on using statistics in astrophysics. Special thanks to the anonymous referee whose comments greatly improved this manuscript.

This work is supported by an NSF Astronomy and Astrophysics Postdoctoral Fellowship under award AST-1501418.

Kepler was competitively selected as the tenth Discovery mission. Funding for this mission is provided by NASA's Science Mission Directorate.

\begin{deluxetable*}{lccccccccc}
\tablecolumns{10}
\tablecaption{Summary statistics for the final 4041 flare star sample. Masses are determined from isochrone fits using the $g-K$ color provided in the KIC, as described in \S\ref{sec:find3}. Rotation periods come from \citet{mcquillan2014}. $\alpha$ and $\beta$ are the power-law fit coefficients to the FFDs. The entire table is provided in machine readable format online.}
\tablehead{
	\colhead{KID \#}&
	\colhead{$g-i$} &
	\colhead{Mass} &
	\colhead{$P_{rot}$} &
	\colhead{$N_{flares}$} &
	\colhead{$N_{flares}$} &
	\colhead{$L_{fl}/L_{Kp}$} &
	\colhead{$\sigma(L_{fl}/L_{Kp})$} &
	\colhead{$\alpha$} &
	\colhead{$\beta$} \\
	\colhead{} &
	\colhead{(mag)} &
	\colhead{$(M_\odot)$} &
	\colhead{(days)} &
	\colhead{} &
	\colhead{$(E>E_{68})$} &
	\colhead{} &
	\colhead{} &
	\colhead{} &
	\colhead{}
	}
\startdata
10000490 &\ldots & 1.38 & \ldots & 241 & 45 & 4.31\e{-5} & 1.48\e{-7}  & 18.83 &	-0.55\\
10001145 & 0.013 & 1.60 & \ldots & 271 & 61 & 5.18\e{-5} & 1.43\e{-8}  & 48.85 &	-1.40\\
10001154 & 1.404 & 0.72 & \ldots & 118 & 115 & 1.43\e{-5} & 1.64\e{-8}& 17.34 &	-0.56\\
10001167 & 1.151 & 0.77 & \ldots & 147 & 131 & 7.24\e{-5} & 2.67\e{-8}& 12.79 &	-0.41\\
10002792 & 1.393 & 0.73 & 1.165 & 225 & 210 & 4.10\e{-4} & 3.38\e{-7}& 16.81 &	-0.52\\
10002897 & 0.079 & 1.49 & \ldots & 155 & 146 & 6.35\e{-5} & 4.56\e{-7}& 11.18 &	-0.28\\
10004510 & 1.449 & 0.71 & 1.373 & 142 & 128 & 7.23\e{-4} & 2.40\e{-7}& 13.82 &	-0.43\\
10004660 & -0.252 & 1.88 & \ldots & 135 & 68 & 2.19\e{-5} & 6.49\e{-9}& 64.71 &	-1.83\\
10005966 & 1.318 & 0.74 & \ldots & 175 & 143 & 3.26\e{-5} & 1.27\e{-8}& 25.21 &	-0.79\\
10006158 & 1.184 & 0.77 & \ldots & 279 & 237 & 5.61\e{-5} & 1.40\e{-8}& 27.63 &	-0.85
\enddata
\label{tbl:datatable}
\end{deluxetable*}

\end{document}